\documentclass[reprint, amssymb, amsmath, pre]{revtex4-1}
\usepackage{bm}
\usepackage{float}
\usepackage{booktabs,tabularx,array}
\usepackage{graphicx,color}
\usepackage[colorlinks=true,allcolors=blue]{hyperref}%
\usepackage{enumerate}
\usepackage{ulem}
\usepackage{harpoon}
\usepackage{amsmath}
\usepackage{gensymb}

\definecolor{mmcolor}{rgb}{0.8, 0.0, 0.2}

\definecolor{ctcolor}{rgb}{0.2, 0.0, 0.8}

\definecolor{ref1color}{rgb}{0.0, 0.0, 1.0}
\definecolor{ref2color}{rgb}{1.0, 0.0, 0.0}

\begin{document}
  
\title{Stiffening of under-constrained spring networks under isotropic strain}
\author{Cheng-Tai Lee}
\author{Matthias Merkel}
\affiliation{CNRS, Centre de Physique Théorique (CPT, UMR 7332), Turing Center for Living Systems, Aix Marseille Univ, Universit\'e de Toulon, Marseille, France}
\begin{abstract}%
Disordered spring networks are a useful paradigm to examine macroscopic mechanical properties of amorphous materials.
Here, we study the elastic behavior of under-constrained spring networks, i.e.\ networks with more degrees of freedom than springs. While such networks are usually floppy, they can be rigidified by applying external strain.
Recently, an analytical formalism has been developed to predict the mechanical network properties close to this rigidity transition.
Here we numerically show that these predictions apply to many different classes of spring networks, including phantom triangular, Delaunay, Voronoi, and honeycomb networks.
The analytical predictions further imply that the shear modulus $G$ scales linearly with isotropic stress $T$ close to the rigidity transition; however, this seems to be at odds with recent numerical studies suggesting an exponent between $G$ and $T$ that is smaller than one for some network classes.
Using increased numerical precision and shear stabilization, we demonstrate here that 
close to the transition
linear scaling, $G\sim T$, holds independent of the network class.
Finally, we show that our results are not or only weakly affected by finite-size effects, depending on the network class.
\end{abstract}
\maketitle

\section*{Introduction}

Understanding macroscopic rigidity and how it depends on the microscopic structure in amorphous materials such as fibrous networks, glasses, jammed colloids, and granular materials has been a long-standing challenge in the field. 
While the macroscopic mechanics of crystalline materials can be computed explicitly by exploiting their spatially periodic microscopic structure, this is not possible for disordered materials.  In particular, upon deformation disordered materials generally display non-affine microscopic displacements, which are hard to predict \cite{Ellenbroek2009c,Silverberg2014,Licup2015,Feng2016,Van_Oosten2016,Sharma2016a,Sharma2016b,Licup2015,Licup2016,Jansen2018,Shivers2019,Arzash2020}.

A classical way to predict the onset of rigidity in many systems is to use Maxwell's constraint counting, which states that rigidity emerges whenever the constraints in a system outnumber its degrees of freedom \cite{Maxwell1864,Calladine1978,Lubensky2015}.
In systems with pair interactions, this is equivalent to comparing the average connectivity $z$, i.e.\ the average number of pair interactions each particle is involved in, to the number of degrees of freedom per particle, which is given by the dimension of space, $D$. Such systems are predicted to be rigid if $z$ exceeds the isostatic point, $z>z_c:=2D$. In this case the system is called over-constrained. Otherwise, for $z<z_c$, the system is called under-constrained or sub-isostatic, and is predicted to be floppy.

\begin{figure}[t]  
\centering	
    \includegraphics[width=0.9\columnwidth]{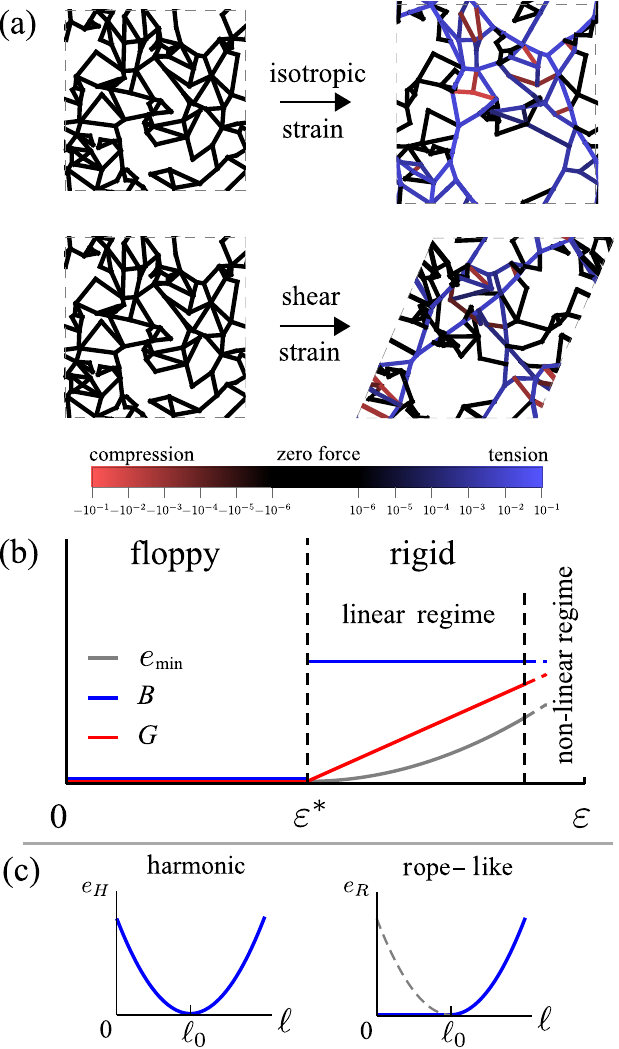} 
	\caption{
	Under-constrained spring networks can be rigidified by the external application of strain. 
	(a) Network configurations and spring tensions for an example network under either isotropic or shear strain.
	(b) Change of minimal network energy $E$, bulk modulus $B$, and shear modulus $G$ when isotropically deforming a network across the rigidity transition, which occurs at the critical strain value $\varepsilon^\ast$.  
	Below $\varepsilon^\ast$, all springs can attain their rest lengths, the system is floppy, and $E=B=G=0$.
	Beyond $\varepsilon^\ast$, springs start to deviate from their rest lengths. According to Ref.~\cite{Merkel2019}, the bulk modulus $B$ shows a discrete jump at $\varepsilon^\ast$, while  $G$ increases linearly and $E$ increases quadratically with the distance from the transition point $\varepsilon^\ast$.
	(c) Two types of spring potentials are used in our simulations, harmonic (left) and rope-like (right). $\ell$ and $\ell_0$ denote spring length and rest length, respectively.
	}
	\label{fig:general concepts}
\end{figure}

While Maxwell's constraint counting predicts under-constrained systems to be floppy, these systems can still be rigidified, either through the application of external strain or the presence of residual stresses \cite{Alexander1998,Wyart2008,Ingber2014,Licup2015,Sharma2016a,Sharma2016b,Merkel2019,Arzash2019,Cui2019,Arzash2020,Damavandi2021}.
As a simple model to study such strain-induced rigidity, we discuss here strain-induced rigidity in athermal, under-constrained disordered spring networks  \cite{Onck2005,Wyart2008,Sheinman2012,Vermeulen2017,Arzash2019,Merkel2019}.
Strain-induced rigidification is illustrated in \autoref{fig:general concepts}a for a network to which isotropic and shear strain has been applied.

The mechanism creating strain-induced rigidity has been discussed in the literature before \cite{Alexander1998,Wyart2008,During2014,Rens2018}.
When approaching the transition from the floppy side, a state of self-stress (SSS) forms right at the transition.
A SSS is a set of tensions that could be put on the springs without any net forces on the nodes. The SSS that appears at the rigidity transition couples to isotropic strain, and using known approaches it can be shown that this induces a jump in the bulk modulus right at the transition (\autoref{fig:general concepts}b) \cite{Lubensky2015,Merkel2019}. 
Meanwhile, the shear modulus shows a continuous transition, whenever the SSS that appears at the transition has no net overlap with shear strain.
Previously, the floppy side of the strain-stiffening transition was discussed in the limit where the springs are infinitely rigid \cite{During2014,Rens2018}.
Here, we are interested in the network mechanics of the rigid side of the transition when spring constants are finite.

Recent work involving one of us proposed a theoretical approach that allows to analytically predict the elastic properties of under-constrained materials close to the strain-induced rigidity transition \cite{Merkel2019}.
This approach is based on a minimal-length function that formalizes the relationship between spring lengths and the applied global strain.
This minimal-length function both reflects the critical point where the network starts to rigidify and allows to predict the elastic network properties in the rigid regime.
In Ref.~\cite{Merkel2019}, this approach was numerically verified both on models for disordered cellular materials and for packing-derived spring networks.
However, it has never been explicitly tested for other classes of under-constrained spring networks.

The approach in Ref.~\cite{Merkel2019} allows to predict the behavior of the elastic moduli close to the transition, where the bulk modulus $B$ shows a discontinuity, while the shear modulus $G$ increases linearly with isotropic strain $\varepsilon$ (\autoref{fig:general concepts}b).
One can show that as a consequence of both, one would expect the shear modulus $G$ to linearly increase also with isotropic stress $T$ close to the transition.
This is also consistent with earlier work on stress-induced rigidity  \cite{Alexander1998,Licup2015,Merkel2018,Lerner2019}.
However, more recent numerical work \cite{Arzash2019} on under-constrained disordered spring networks suggested that the value for the scaling exponent between $G$ and $T$ can differ from one, depending on the class of network studied.
The reason for this deviation from the analytical predictions is so far unclear.
Other recent work proposed that the numerical results in Ref.~\cite{Merkel2019} could potentially be affected by finite-size effects caused by a diverging length scale when shearing the system \cite{Arzash2020}.
Could similar finite-size effects be the reason for this contradiction between predicted and numerically obtained exponents between $G$ and $T$?

Here, we numerically test the predictions from Ref.~\cite{Merkel2019} on several different classes of athermal spring networks. These include phantom triangular, Delaunay, Voronoi, and honeycomb networks, where we study two types of spring potentials, harmonic and rope-like (\autoref{fig:general concepts}c).
In the following, we first summarize the analytical approach from Ref.~\cite{Merkel2019} in \autoref{sec:analytics}. 
We then test the analytical predictions on the four  different network classes in \autoref{sec:numerics}, and show that they follow the predicted behavior (\autoref{sec:floppy-rigid transition and linear relation}).
In \autoref{sec:z-scaling}, we furthermore show that the scaling behavior of the coefficients appearing in the minimal-length function with connectivity $z$ depends on the network class.
We then numerically explore the scaling behavior of the shear modulus $G$ over isotropic stress $T$ with increased numerical precision and find a scaling exponent of one, independent of network class (\autoref{sec:shear modulus scales linearly}). 
Finally, we show that depending on the network class, there is no or a weak system-size dependence affecting these results (\autoref{sec:not from finite-size effect}).

\section{Theoretical predictions}
\label{sec:analytics}
We start by summarizing the approach of Ref.~\cite{Merkel2019}, which allows to predict the elastic properties of under-constrained spring networks close to the rigidity transition. 

In general, the formalism of Ref.~\cite{Merkel2019} applies to any disordered Hookean spring network of $N$ springs, where each spring $i$ has a different spring constant $k_i$ and rest length $\ell_{0i}$. The energy of such a network is:
\begin{equation}\label{eq:E_general}
e=\sum_{i=1}^{N}{k_i\left(\ell_{i} - \ell_{0i} \right)^2},
\end{equation}
where $\ell_i$ is the length of spring $i$.
The springs are connected at movable nodes, around which they can freely rotate. While the approach can be applied largely independently of the precise boundary conditions, we focus here on periodic boundary conditions with fixed system size. 
Unless stated otherwise, we use dimensionless quantities, where the length unit is $L_c:=(V/N)^{1/D}$ with $D$ being the dimension of space and $V$ the system volume.
We define the energy unit such that $(\sum_i k_i)/N=1$.
Using dimensionless lengths will later allow us to describe the effect of isotropic strain (\autoref{sec:effect of bulk strain}).

Here, to explain just the key ideas of the approach, we focus for simplicity on the special case of a network with homogeneous spring constants $k_i=1$ and rest lengths $\ell_{0i}=\ell_0$:
\begin{equation}\label{eq:E_homogeneous}
e=\sum_{i=1}^{N} \left(\ell_{i} - \ell_{0} \right)^2.
\end{equation}
The behavior of networks with heterogeneous spring properties can be predicted by formally mapping them onto Eq.~\eqref{eq:E_homogeneous} as discussed in appendix~\ref{sec:heterogeneous}.

The elastic properties of disordered networks are in general difficult to predict analytically.
Formally, these elastic properties can be computed from derivatives of a minimal energy function $e_\mathrm{min}(\varepsilon,\gamma)$ with respect to external isotropic strain $\varepsilon$ or shear strain $\gamma$. This function corresponds to the minimized system energy $e(\lbrace\bm{r}_n\rbrace,\varepsilon,\gamma)$ with respect to the node positions $\bm{r}_n$ at constant strain variables $\varepsilon,\gamma$.
However, applying strain to a disordered network generally induces non-affine displacements of the node positions, which are typically hard to predict without numerical energy minimization.
To nevertheless make non-trivial predictions about the elastic network properties, Ref.~\cite{Merkel2019} introduced a different approach.
Instead of explicitly following the node motion, progress can already be made by focusing on the relation between spring lengths and external strain.

Note that while we focus in this section on harmonic springs, the formalism can also be applied to networks with rope-like pair interactions (\autoref{fig:general concepts}c). 
This is because a rope-like pair interaction can be perfectly mimicked by a chain of two or more harmonic springs \cite{Merkel2019} (see also appendix~\ref{sec:parameter extraction}).

\subsection{Key idea}\label{sec:key_idea}
To obtain an explicit expression for $e_\mathrm{min}$ in terms of external strain, we first transform the expression in Eq.~\eqref{eq:E_homogeneous} into a sum of two squares:
\begin{equation}\label{eq:E two terms}
e = N \left[ \left( \bar{\ell}- \ell_0 \right)^2 + \sigma_{\ell}^2 \right].
\end{equation} 
Here, $\bar{\ell}=(\sum_i{\ell_i})/N$ and $\sigma_\ell^2=(\sum_i{[\ell_i-\bar{\ell}]^2})/N$ are average and variance of the spring lengths, respectively.

The expression in Eq.~\eqref{eq:E two terms} allows us to more conveniently discuss the minimal network energy $e_\mathrm{min}$ and its behavior once we strain the system.
Because $e$ is the sum of two squares, an energy minimum is attained whenever both $|\bar{\ell}-\ell_0|$ and $\sigma_{\ell}$ are as small as possible.
There are two possibilities.
First, if there is a set of node positions such that both squares can simultaneously attain zero, then the minimal energy is zero $e_\mathrm{min}=0$.  Because elastic stresses and moduli correspond to derivatives of $e_\mathrm{min}$, the system is floppy in this parameter regime.
Second, there might be \textit{no} set of node positions such that both terms $|\bar{\ell}-\ell_0|$ and $\sigma_{\ell}$ can simultaneously vanish. In this regime, the system is typically rigid.

To access the value of $e_\mathrm{min}$ in the rigid regime, we need to understand how the system compromises between minimizing $|\bar{\ell}-\ell_0|$ and $\sigma_{\ell}$ in Eq.~\eqref{eq:E two terms}.
To this end, we first need a way to express which combinations of $\bar{\ell}$ and $\sigma_\ell$ are geometrically possible.  
As shown in Ref.~\cite{Merkel2019}, this can be done using a minimal-length function $\bar{\ell}_\mathrm{min}(\sigma_\ell)$, which returns the minimally possible $\bar\ell$ for a given $\sigma_\ell$. In other words, a combination of $\bar{\ell}$ and $\sigma_\ell$ is \textit{geometrically possible} only if:  
\begin{equation}\label{eq:Lmin}
  \bar{\ell}\geq \bar{\ell}_\mathrm{min}(\sigma_\ell).
\end{equation}
For instance, for $\sigma_\ell=0$ it is possible to find only network configurations with $\bar{\ell}\geq \bar{\ell}_\mathrm{min}(\sigma_\ell=0)$. Thus, for $\ell_0\geq \bar{\ell}_\mathrm{min}(\sigma_\ell=0)$ the network will be floppy, because both squares in Eq.~\eqref{eq:E two terms} can simultaneously vanish, which implies that $e_\mathrm{min}$ and its derivatives vanish.
Conversely, for $\ell_0<\bar{\ell}_\mathrm{min}(\sigma_\ell=0)$, the first term in Eq.~\eqref{eq:E two terms} can not vanish with $\sigma_\ell=0$, because only configurations with $\bar\ell\geq\bar{\ell}_\mathrm{min}(\sigma_\ell=0)>\ell_0$ are possible.
Thus, $\ell_0^\ast:=\bar{\ell}_\mathrm{min}(\sigma_\ell=0)$ is the transition point between floppy and rigid regime.

In general, the precise functional form of $\bar{\ell}_\mathrm{min}(\sigma_\ell)$ depends on the network structure.
However, we showed in Ref.~\cite{Merkel2019} that to first order in $\sigma_\ell$ it can be expanded as
\begin{equation}\label{eq:Lmin expansion}
  \bar{\ell}_\mathrm{min}(\sigma_\ell) = \ell_0^\ast - a_\ell\sigma_\ell,
\end{equation}
where $\ell_0^\ast$ and $a_\ell$ are constants that encode the network structure. 
Eq.~\eqref{eq:Lmin expansion} holds in the limit of small $\sigma_\ell$, which means that the system is close to the transition point, where $\sigma_\ell=0$.
Note that Eq.~\eqref{eq:Lmin expansion} is closely related to the SSS that is created at the transition, where $a_
\ell$ is the coefficient of variation (standard deviation over mean) of the SSS components.
We expect that deriving expressions for $\ell_0^\ast$ and $a_\ell$ from first principles is very hard for disordered networks. Besides some exceptions, $\ell_0^\ast$ and $a_\ell$ will need to be determined numerically.

To derive an expression for the minimal energy $e_\mathrm{min}$ in the solid regime, we combine two parts: the energy in Eq.~\eqref{eq:E two terms} and the condition of geometrically possible combinations $(\bar{\ell},\sigma_\ell)$ in Eqs.~\eqref{eq:Lmin} and \eqref{eq:Lmin expansion}. First, Eq.~\eqref{eq:Lmin} implies that for fixed $\sigma_\ell$, the energy in Eq.~\eqref{eq:E two terms} is minimized when $\bar{\ell}=\bar{\ell}_\mathrm{min}(\sigma_\ell)$. Combining this with Eq.~\eqref{eq:Lmin expansion}, insertion into Eq.~\eqref{eq:E two terms}, and minimization with respect to $\sigma_\ell$, yields:
\begin{equation}\label{eq:E explicit}
e_\mathrm{min} = \frac{N}{1+a_\ell^2} \left( \ell_0^\ast- \ell_0 \right)^2.
\end{equation}
This expression only depends on the spring number $N$, the spring constant $k$, the rest length $\ell_0$, and the two parameters $\ell_0^\ast$ and $a_\ell$ that encode the network structure.
Note that from Eq.~\eqref{eq:E explicit} we see that the system energy is that of a single effective spring with rest length $\ell_0$.

\subsection{Simple example network}\label{sec:simple example}
\begin{figure}[t]  
\centering	\includegraphics[width=1\columnwidth]{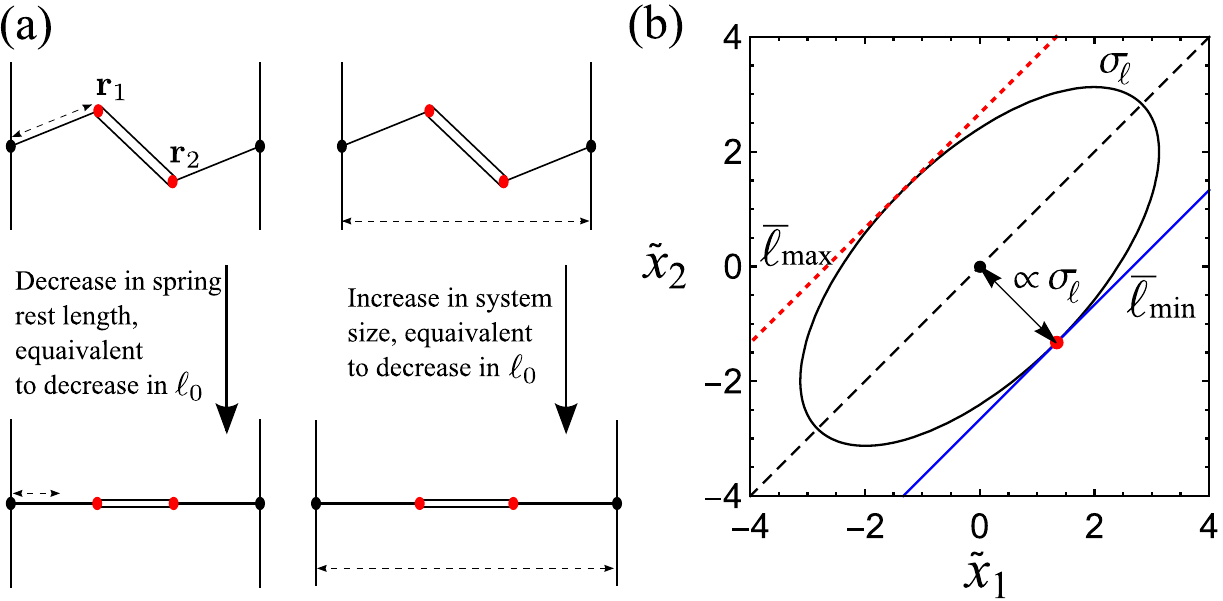}
	\caption{
	Illustration of the analytical formalism using a 4-spring example network.
	(a) The network can be rigidified by either decreasing the dimensional spring rest length or increasing the system size, both of which has the effect of decreasing the dimensionless parameter $\ell_0$. Black and red dots indicate fixed and movable nodes, respectively. 
	(b) Dependence of average $\bar{\ell}$ and standard deviation $\sigma_\ell$ of the four spring lengths on the internal node positions $\bm{r}_{n}=(x_n, y_n)$ with $i\in\lbrace 1, 2\rbrace$ and $y_1=y_2=0$.  The axes are $\tilde x_1 = x_1 + 1/6$ and $\tilde x_2 = x_2- 1/6$.
	Curves of constant $\bar{\ell}$ are diagonal lines (with increasing $\bar{\ell}$: blue solid, black dashed, red dotted lines), while curves of constant $\sigma_\ell$ are ellipses.
        The configuration of minimal $\bar{\ell}$ for given $\sigma_\ell$ is indicated by the red dot.
        Because the linear size of the ellipse scales with $\sigma_\ell$, the minimal $\bar{\ell}$ for given $\sigma_\ell$ decreases linearly with $\sigma_\ell$.
	}
	\label{fig:spEx}
\end{figure}

To illustrate the ideas of the previous section, we discuss a simple example network (\autoref{fig:spEx}a left). The network consists of four springs with equal dimensionless spring constants $k=1$ and rest lengths $\ell_0$. Two of the springs are connected to fixed points (black dots) located at positions $(-1/2,0)$ and $(1/2,0)$, respectively. The two internal nodes (red dots) at positions $\bm{r}_n$ with $n=1,2$ are movable. We will use the ideas of the previous section to derive an expression for the minimal energy $e_\mathrm{min}$.

For $\ell_0\geq 1/3$, there are always configurations where all springs can attain their rest lengths $\ell_i=\ell_0$ (\autoref{fig:spEx}a top). This implies that $|\bar{\ell}-\ell_0|=0$ and $\sigma_{\ell}=0$, i.e.\ both terms in Eq.~\eqref{eq:E two terms} can simultaneously vanish.

Conversely, for $\ell_0<1/3$, the springs will be under tension (\autoref{fig:spEx}a bottom). 
Our 4-spring example network is simple enough so that we can explicitly minimize the energy with respect to the inner node positions $\bm{r}_n=(x_n,y_n)$ with $n=1,2$.
This will allow us to first directly test whether the minimal energy has the form predicted by Eq.~\eqref{eq:E explicit} in the previous section.
The energy of our example network is
\begin{equation}
    \begin{aligned}
      e &= \bigg[
      \left(x_1+\frac{1}{2}-\ell_0\right)^2
      +2\Big(x_2-x_1-\ell_0\Big)^2 \\
      &\qquad\quad+\left(\frac{1}{2}-x_2-\ell_0\right)^2
      \bigg].
    \end{aligned}
\end{equation}
Here, to simplify the following discussion, we have set $y_1=y_2=0$.
The energy $e$ has a global minimum at $x_1=-(1+2\ell_0)/10, x_2=(1+2\ell_0)/10$, where its value is
\begin{equation}\label{eq:energy example final}
    e_\mathrm{min} = \frac{2}{5}\big(1 - 3\ell_0\big)^2.
\end{equation}
This expression is indeed of the predicted form Eq.~\eqref{eq:E explicit}.

We now demonstrate how the minimal energy $e_\mathrm{min}$ can instead be obtained using the ideas of the previous section.
We first discuss which pairs of $\bar{\ell}$ and $\sigma_\ell$ are geometrically possible. To this end, we express $\bar{\ell}$ and $\sigma_\ell$ in terms of the internal degrees of freedom $x_1,\ x_2$:
\begin{align}
\bar{\ell} &= \frac{1}{3} + \frac{1}{4}\big(\tilde{x}_2-\tilde{x}_1\big), \label{eq:Lavg in toy model}\\
\sigma_\ell^2 &= \frac{1}{16}\Big(11\tilde{x}_1^2-14 \tilde{x}_1 \tilde{x}_2 +11\tilde{x}_2^2\Big),\label{eq:sigma in toy model}
\end{align}
where we defined $\tilde{x}_1=x_1+1/6$ and $\tilde{x}_2=x_2-1/6$.
Both Eqs.~\eqref{eq:Lavg in toy model} and \eqref{eq:sigma in toy model} are illustrated in \autoref{fig:spEx}b.
Curves of constant $\bar{\ell}$ correspond to lines inclined by $45\degree$, where $\bar{\ell}$ increases as $\tilde{x}_1$ decreases and $\tilde{x}_2$ increases.
Meanwhile, curves of constant $\sigma_\ell$ correspond to ellipses centered at $\tilde{x}_1=\tilde{x}_2=0$, whose main axes scale with $\sigma_\ell$ and are oriented at $45\degree$ angles with respect to the $\tilde{x}_1$ and $\tilde{x}_2$ axes \footnote{This is because Eq.~\eqref{eq:sigma in toy model} can be transformed into:
\begin{equation}
\sigma_L^2 = \frac{1}{16}(2u^2+9w^2),
\end{equation}
where $u=\tilde{x}_1+\tilde{x}_2$ and $w=\tilde{x}_1-\tilde{x}_2$.
This is the equation of an ellipse whose main axes are diagonally oriented and scale with $\sigma_\ell$.
}.
Thus, for a given value of $\sigma_\ell$, any combination of $x_1$ and $x_2$ can give rise to values for $\bar{\ell}$ only in an interval between $\bar{\ell}_\mathrm{min}(\sigma_\ell)$ (blue solid line) and $\bar{\ell}_\mathrm{max}(\sigma_\ell)$ (red dashed line).
The upper bound $\bar{\ell}_\mathrm{max}(\sigma_\ell)$ only exists because we set $y_1=y_2=0$ before; without this constraint, $\bar{\ell}$ can become arbitrarily large for a given $\sigma_\ell$ \footnote{That $\bar{\ell}$ can become arbitrarily large for given $\sigma_\ell<1/4$ can be shown explicitly by considering a subset of configurations parameterized by two scalars $w$ and $h$ as $\bm{r}_1=(-w/2,h)$ and $\bm{r}_2=(w/2,h)$. Then one can show that the choice
\begin{equation*}
  w(\sigma_\ell,h) = -\frac{1}{3}(1+8\sigma_\ell) + \frac{2}{3}\sqrt{(1+2\sigma_\ell)^2 + 3h^2}
\end{equation*}
leads to the correct value for the standard deviation of the spring lengths $\sigma_\ell$. Moreover, one can show that for this choice, the relation
\begin{equation*}
  \bar{\ell}(\sigma_\ell,h) = \sigma_\ell + w(\sigma_\ell,h)
\end{equation*}
holds, and that $\bar{\ell}(\sigma_L\ell,h=0)=\bar{\ell}_\mathrm{min}(\sigma_\ell)$ with $\bar{\ell}_\mathrm{min}$ given by Eq.~\eqref{eq:linear Lmin eq}. Finally, for fixed $\sigma_\ell$, the function $\bar{\ell}(\sigma_\ell,h)$ increases monotonically with $h$ without upper bound.}.
Meanwhile, the lower bound $\bar{\ell}_\mathrm{min}(\sigma_\ell)$ decreases linearly with the distance between origin (black dot) and the intersection point (red dot) in \autoref{fig:spEx}b, which is proportional to $\sigma_\ell$. 
As a consequence, using Eqs.~\eqref{eq:Lavg in toy model} and \eqref{eq:sigma in toy model}:
\begin{equation}\label{eq:linear Lmin eq}
    \bar{\ell}_\mathrm{min}(\sigma_\ell) = \frac{1}{3} - \frac{\sigma_\ell}{3}.
\end{equation}
This is of the form of Eq.~\eqref{eq:Lmin expansion}, where we identify $\ell_0^\ast=1/3$ and $a_\ell=1/3$. Inserting this into Eq.~\eqref{eq:E explicit}, we obtain indeed Eq.~\eqref{eq:energy example final}.

In our discussion here we included the internal degrees of freedom $x_1, x_2$ to demonstrate their connection to geometrically possible combinations of $\bar{\ell}$ and $\sigma_\ell$, and to obtain explicit values for $\ell_0^\ast$ and $a_\ell$.  In general, however, the approach from Ref.~\cite{Merkel2019} does not require a discussion of internal degrees of freedom. Equations~\eqref{eq:E two terms}--\eqref{eq:Lmin expansion} are sufficient to understand the overall system behavior close to the rigidity transition, unless one wants to derive the values of the coefficients $\ell_0^\ast$ and $a_\ell$ from first principles.

\subsection{Effect of isotropic strain}\label{sec:effect of bulk strain}
We now discuss how the effect of isotropic strain $\varepsilon$ is incorporated into the formalism. The 4-spring system in \autoref{fig:spEx}a transitions from floppy to rigid when decreasing the dimensionless parameter $\ell_0$.
Such a decrease in $\ell_0$ can correspond either to a decrease in the \textit{dimensional} spring rest length while keeping the system size constant (\autoref{fig:spEx}a left), or to an increase in system size while keeping the dimensional rest length constant (\autoref{fig:spEx}a right). 
Thus, $\ell_0$ is a control parameter combining both dimensional spring rest length and isotropic strain.

Let us consider simulations where the dimensional spring rest length $L_0$ is kept constant, but the system size $V$ is changing. In this case, the combined control parameter $\ell_0$ encodes isotropic strain. We define (linear) isotropic strain as $\varepsilon := (V/V_\mathrm{ref})^{1/D} - 1$, where the $V_\mathrm{ref}$ is the system volume right after creation of the network. From our length non-dimensionalization follows that we can convert between $\ell_0$ and bulk strain $\varepsilon$ via:
\begin{equation} \label{eq:l0 epsilon}
  \ell_0 = \frac{L_0}{1+\varepsilon}\left(\frac{N}{V_\mathrm{ref}}\right)^{1/D}.
\end{equation}
Inserting this equation into Eq.~\eqref{eq:E explicit} provides an explicit expression of the dimensionless system energy on isotropic strain $\varepsilon$.

\subsection{Effect of shear strain}\label{sec:effect of shear strain}
To understand how shear strain enters the formalism, we first note that shearing the system does not change the energy formula Eq.~\eqref{eq:E two terms}.
However, shearing the system will change the set of geometrically possible combinations $(\bar\ell,\sigma_\ell)$.  Thus, shear strain needs to be included as a parameter in the minimal-length function $\bar\ell_\mathrm{min}$. In Ref.~\cite{Merkel2019} this function is Taylor expanded to second order in shear strain, so that Eq.~\eqref{eq:Lmin expansion} becomes most generally:
\begin{equation}\label{eq:lmin with hat shear}
\bar\ell_{\rm{min}}(\sigma_\ell,\hat\gamma) = \hat\ell_0^\ast-a_{\ell}\sigma_{\ell}+b_1\hat\gamma+b\hat\gamma^2.
\end{equation}
For later compactness of notation, here we also substituted the notation of parameter $\ell_0^\ast$ by $\hat\ell_0^\ast$.

Note that the linear order term in $\hat\gamma$ appears only because disordered systems with a finite size generally display a small but finite anisotropy.
Equation~\eqref{eq:lmin with hat shear} can be simplified by removing this anisotropy through defining a new shear variable $\gamma=\hat\gamma-\Delta\gamma_0$, where $\hat\gamma=\Delta\gamma_0$ is defined as the shear where the function $\bar\ell_{\rm{min}}(\sigma_\ell,\hat\gamma)$ is minimal: $\Delta\gamma_0=-b_1/2b$.  Defining $\ell_0^\ast:=\hat\ell_0^\ast -b_1^2/4b$, this leads to the minimal-length function:
\begin{equation}\label{eq:lmin with shear}
\bar\ell_{\rm{min}}(\sigma_\ell,\gamma) = \ell_0^\ast-a_{\ell}\sigma_{\ell}+b\gamma^2.
\end{equation}
The anisotropy $\Delta\gamma_0$ is expected to disappear in the limit of a large network size.

\subsection{Elastic properties near the rigidity transition}\label{sec:mechanic properties}
Substituting Eq.~\eqref{eq:lmin with shear} into Eq.~\eqref{eq:E two terms} and minimizing with respect to $\sigma_{\ell}$, we obtain the following explicit energy expression in terms of the control parameters $\ell_0$ and $\gamma$:
\begin{equation}\label{eq:e with shear}
    e_{\mathrm{min}}(\ell_0,\gamma)=\frac{N}{1+a_{\ell}^2}\Big( \ell_0^*-\ell_0+b\gamma^2\Big)^2.  
\end{equation}
Derivatives of this expression with respect to $\ell_0$ (which is related to isotropic strain $\varepsilon$ via Eq.~\eqref{eq:l0 epsilon}) and shear strain $\gamma$ allow to derive the following quantities, here for the 2D case \cite{Merkel2019}:
\begin{align}
T &=  \frac{\ell_0^\ast}{1+a_{\ell}^2}\Big(\ell_0^\ast-\ell_0+b\gamma^2\Big),\label{eq:P prediction}\\
\sigma &=  \frac{4b\gamma}{1+a_{\ell}^2}\Big(\ell_0^\ast-\ell_0+b\gamma^2\Big), \label{eq:shear stress prediction}\\
\Delta B &=\frac{\left(\ell_0^\ast\right)^2}{2\left(1+a_{\ell}^2\right)}\ \ \text{at $\gamma=0$},\label{eq:B prediction} \\
G &= \frac{4 b}{1+a_{\ell}^2} \Big(\ell_0^\ast-\ell_0+3b\gamma^2\Big).\label{eq:G prediction}
\end{align}
Here, $T$, $\sigma$, $\Delta B$ and $G$ are isotropic stress, shear stress, bulk modulus discontinuity, and shear modulus, respectively.  These formulas hold close to the rigidity transition in the region where Eq.~\eqref{eq:lmin with shear} is accurate.
As shown by Eqs.~\eqref{eq:P prediction}--\eqref{eq:G prediction}, the three parameters $\ell_0^\ast$, $a_\ell$, and $b$ fully describe the macroscopic elastic behavior in this regime.  

\section{Numerical results}
\label{sec:numerics}
While in Ref.~\cite{Merkel2019} the analytical predictions in Eqs.~\eqref{eq:P prediction}--\eqref{eq:G prediction} were numerically tested on packing-derived networks only, we test these predictions here on a set of additional network classes.  These include phantom triangular and Delaunay networks (both with varying connectivity $z$), as well as honeycomb and Voronoi networks (which both have fixed connectivity $z=3$).  We probe the elastic properties of these networks under isotropic (i.e.\ bulk) strain.

\begin{figure*}[t]  
\centering	\includegraphics[width=2\columnwidth]{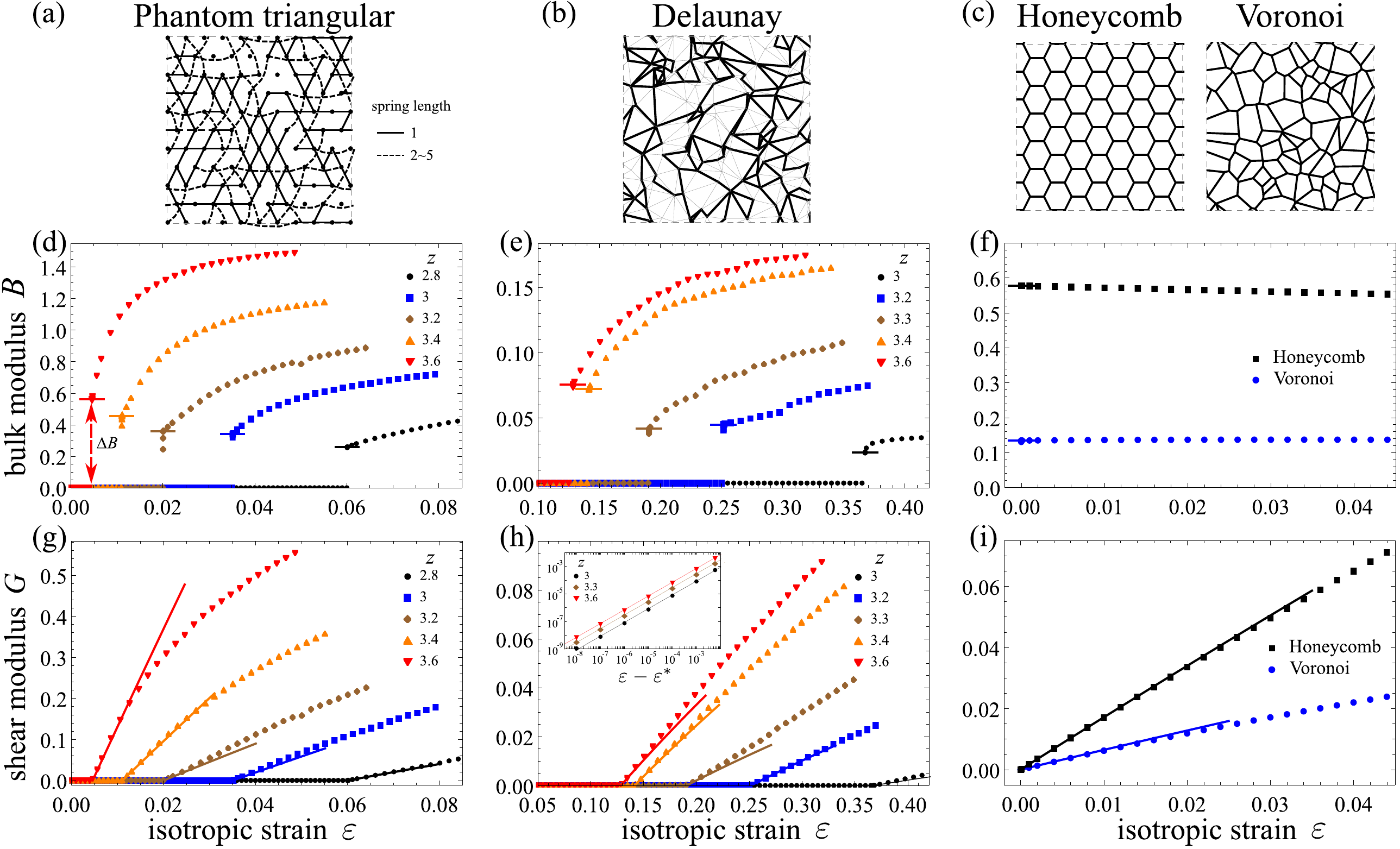}
	\caption{
	Behavior of bulk and shear moduli across the transition for four different classes of networks and comparison to analytical predictions.
	(a)-(c) Sketches of network structures of phantom triangular (a, $z=3.2$, $W=10$), Delaunay (b, $z=3.2$, $W=12$), Voronoi (c, $z=3$, $W=12$) and honeycomb (c, $z=3$, $W=12$) networks. In (a), the solid lines are springs of length one, while the dashed ones are phantom springs, i.e.\ springs that cross one or more nodes without being connected to them. We show phantom springs, which are actually straight, slightly curved only for better visualization. In (b), gray lines indicate the removed springs from the initial full Delaunay network, leaving the black springs in the actual network. 
	(d)-(i) Numerically obtained bulk modulus $B$ and shear modulus $G$ at $\gamma=0$ against increasing isotropic strain $\varepsilon$ for different network classes: phantom triangular (d, g, $W=40$) and Delaunay (e, h, $W=20$) networks with variable connectivity $z$, as well as Voronoi (f, i, $W=70$) and honeycomb (f, i, $W=60$). Here, we use harmonic spring potentials, and we shear stabilized the networks before the $\varepsilon$ sweeps (appendix~\ref{sec:details sweeps}). The discontinuity $\Delta B$ in the bulk modulus at the transition point and linear scaling of $G$ predicted from Eqs.~\eqref{eq:B prediction} and \eqref{eq:G prediction} are indicated as solid bars and solid lines, respectively. (h inset) Log-log plot of $G$ against strain difference $\varepsilon-\varepsilon^\ast$, to resolve the vicinity of the transition point $\varepsilon^\ast$. Symbols are numerical data and lines are analytical predictions.
	}
	\label{fig:floppy-rigid transition}
\end{figure*}

\subsection{Network generation and energy minimization}
\label{sec:network generation and minimization}
Networks of freely hinging nodes are created in a periodic box following existing protocols \cite{Broedersz2011b,Arzash2019} (details in appendix~\ref{sec:network generation}).
We probe the system by varying isotropic strain $\varepsilon$. Each time, we first use bisection to detect the transition point $\varepsilon^\ast$, before we carry out exponential and/or linear sweeps in isotropic strain $\varepsilon$ (details in appendix~\ref{sec:details sweeps}).
To ensure high precision in our energy minimization, we use an optimized conjugate gradient scheme that allows to reduce the average residual force per degree of freedom to less than $10^{-12}$ \cite{Merkel2019}.

Right after creation, where $\hat\gamma=0$, the disordered networks will generally display an anisotropy. To remove this anisotropy, we need to shear the system to the state $\hat\gamma=\Delta\gamma_0$ (i.e.\ $\gamma=0$, see \autoref{sec:effect of shear strain}). At this point, according to Eq.~\eqref{eq:shear stress prediction}, shear stress vanishes, $\sigma=0$. 
Thus, the anisotropy in the networks can be numerically removed using shear stabilization \cite{Dagois-Bohy2012}.  
Shear stabilization means that shear strain is treated as an additional degree of freedom during the energy minimization. 
Unless stated otherwise, we always apply this method during the bisection phase to search for the transition point, so that our system right after the bisection phase is at $(\varepsilon,\gamma)=(\varepsilon^\ast,0)$. During the subsequent $\varepsilon$ sweeps, we keep shear strain $\gamma$ fixed (details in appendix~\ref{sec:details sweeps}).

\begin{figure*}[t]  
\centering	\includegraphics[width=2\columnwidth]{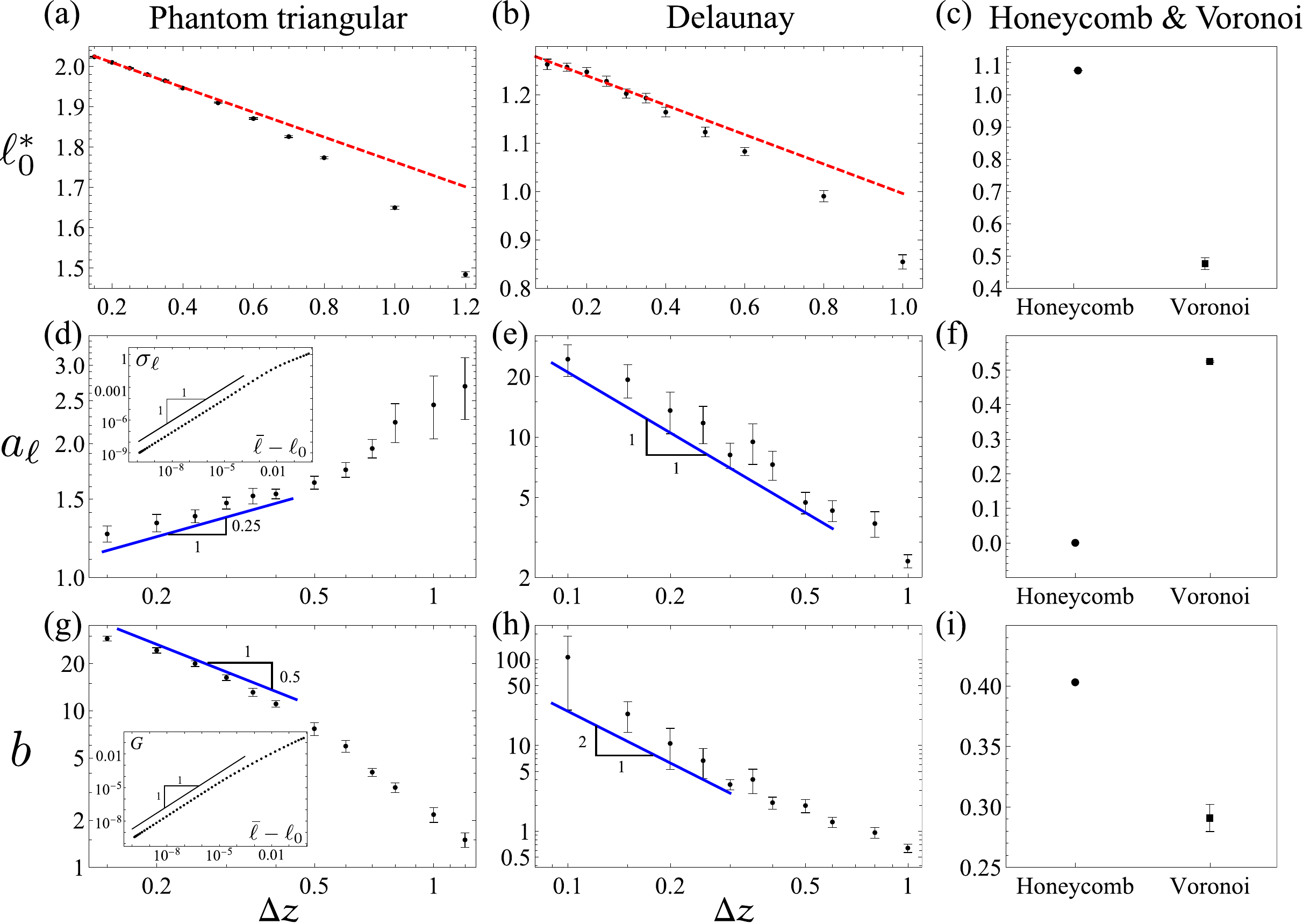}
\caption{Scaling of the parameters $\ell_0^\ast$, $a_{\ell}$ and $b$ with network connectivity $z$ for phantom triangular (system size $W=40$), Delaunay ($W=20$), Voronoi ($W=70$; $z=3$) and honeycomb ($W=60$; $z=3$) networks with harmonic spring potentials. The parameters $\ell_0^\ast$, $a_{\ell}$ and $b$ are extracted for each network realization according to the protocols in \autoref{sec:floppy-rigid transition and linear relation} and appendix~\ref{sec:parameter extraction}.
For instance, (d inset) $a_{\ell}$ is extracted using a linear fit of $\sigma_\ell$ over $\bar\ell-\ell_0$, and (g inset) $b$ is extracted using a linear fit of $G$ over $\bar\ell-\ell_0$.
Error bars in all panels indicate the standard error of the mean. We use the first 5 data points away from $\Delta z=0$ to fit $\ell_0^\ast$ and obtain $2.07-0.31\Delta z$ for phantom triangular and $1.30-0.30\Delta z$ for Delaunay networks (red dashed lines in panels (a) and (b)). For the honeycomb network, the numerically obtained value for $\ell_0^\ast$ is consistent with its theoretical value $\ell_0^\ast=2^{1/2}/3^{1/4} \approx 1.0745$, and the parameter $a_{\ell}$ is exactly zero, because $\sigma_{\ell}=0$ due to symmetry.}
	\label{fig:z-scaling}
\end{figure*}

\subsection{Elastic moduli close to the transition}
\label{sec:floppy-rigid transition and linear relation}
To numerically characterize the nature of the transition, we first carry out a combination of exponential and linear sweeps around the transition point $\varepsilon^\ast$ (details in appendix~\ref{sec:details sweeps}).
In \autoref{fig:floppy-rigid transition}, we plot bulk modulus $B$ and shear modulus $G$ against isotropic strain $\varepsilon$ for single network realizations with varying connectivity $z$, where we use harmonic spring potentials.

At the transition, all networks show a discontinuity $\Delta B$ in the bulk modulus, while the transition is continuous in the shear modulus $G$. This is qualitatively consistent with our analytical predictions (\autoref{sec:mechanic properties}) and the behavior of packing-derived networks \cite{Merkel2019}.
We also observe that for both phantom triangular and Delaunay networks the transition point $\varepsilon^\ast$ decreases with the average connectivity $z$. 

To compare these data to the prediction for the bulk modulus discontinuity $\Delta B$ according to Eq.~\eqref{eq:B prediction}, we need the values of $\ell_0^\ast$ and $a_\ell$ for our simulations.  To extract $\ell_0^\ast$, we insert the transition point strain value $\varepsilon^\ast$ into Eq.~\eqref{eq:l0 epsilon}.  To extract $a_\ell$, we plot $\sigma_\ell$ over $\bar\ell-\ell_0$ (inset of \autoref{fig:z-scaling}d) and perform a linear fit whose slope is $a_\ell$ for small $\bar\ell-\ell_0$ (see appendix~\ref{sec:parameter extraction}). Note that for symmetry reasons, the honeycomb lattice has $\sigma_\ell=0$ and thus $a_\ell=0$.  The resulting predictions for the bulk modulus discontinuities $\Delta B$ are respectively indicated as horizontal bars in \autoref{fig:floppy-rigid transition}d-f.
Indeed, our predictions match well the discontinuities present in the simulation data for all four network classes and all connectivities $z$.

Some of the data points right at the transition fall below the analytically predicted value for $\Delta B$.
These deviations occur in our data for the strain values closest to the transition point, while $\Delta B$ values of the same network at similar strain values match closely with our analytical prediction. These deviations are likely due to numerical residues, an effect that we observed before \cite{Merkel2019}. 

For the shear modulus, Eq.~\eqref{eq:G prediction} predicts a continuous transition with a linear scaling $G\sim(\ell_0^\ast-\ell_0)$.  Using Eq.~\eqref{eq:l0 epsilon}, this implies also a linear scaling $G\sim\Delta\varepsilon:=\varepsilon-\varepsilon^\ast$ to lowest order in $\Delta\varepsilon$.  Indeed, this is what we observed close to the transition (inset of \autoref{fig:z-scaling}g). We indicate this linear scaling also in \autoref{fig:floppy-rigid transition}g-i. Note that for larger $\Delta\varepsilon$, non-linearities in $\bar\ell_\mathrm{min}$ and in Eq.~\eqref{eq:l0 epsilon} create deviations from this prediction.

Note that both honeycomb and Voronoi networks have their transition points at $\varepsilon^\ast=0$.  This means that these networks have a SSS already right at creation. While this is clearly the case for the honeycomb lattice, we show in appendix~\ref{sec:Voronoi self-stress} that it is also true for any Voronoi network. 

Taken together, the elastic properties of the system close to the transition, such as the transition point $\varepsilon^\ast$, the magnitude of the discontinuity $\Delta B$ in the bulk modulus, and the linear scaling coefficient for the shear modulus $G$, can be predicted from the coefficients $\ell_0^\ast$, $a_\ell$, and $b$.

\begin{figure*}[t]  
\centering	\includegraphics[width=2\columnwidth]{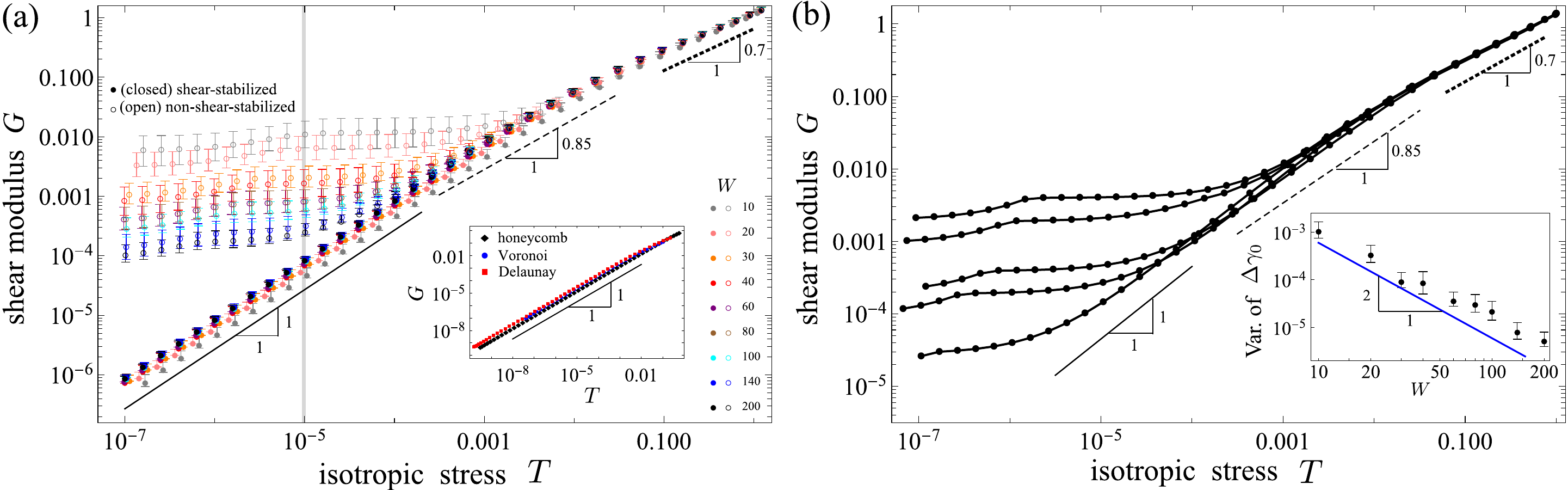}
	\caption{The shear modulus $G$ scales linearly with isotropic stress $T$ for shear-stabilized networks. Without shear stabilization an additional plateau appears at small $T$. 
	(a) Dependence of $G$ on isotropic stress $T$ for phantom triangular networks with $z=3.2$ and rope-like spring potentials for varying system size $W$, where we either use shear stabilization (closed circles) or not (open circles). All shear-stabilized networks show a linear scaling close to the transition point (at small $T$). The gray vertical line indicates the lowest $T$ value probed in previous work  \cite{Arzash2019}. 
	(a inset) The linear scaling in $G(T)$ also appears for shear-stabilized honeycomb ($W=60$), Voronoi ($W=70$) and Delaunay ($W=20$) networks. Error bars in panel a \& inset indicate the standard error of the mean.
	(b) The $G(T)$ curves for individual phantom triangular networks with $z=3.2$ and $W=40$ without shear stabilization also exhibit a plateau, confirming that the plateaus in panel a are not due to an averaging effect. 
	(b inset) 
    The variance of the network anisotropy $\Delta \gamma_0$ (defined below Eq.~\eqref{eq:lmin with hat shear}) across different randomly generated networks for a given system size $W$ scales inversely proportional with $W^2$.
	}
	\label{fig:G_P}
\end{figure*}
\subsection{Scaling of $\ell_0^\ast$, $a_{\ell}$ and $b$ with connectivity $z$}\label{sec:z-scaling}
In \autoref{fig:z-scaling}, we plot the parameters $\ell_0^\ast$, $a_{\ell}$ and $b$ for phantom triangular, Delaunay, Voronoi, and honeycomb networks with harmonic spring potentials. For phantom triangular and Delaunay networks, we show the dependency on the connectivity $z$. For the disordered networks (i.e.\ phantom triangular, Delaunay, and Voronoi) we average each time over 50 random realizations.

In both phantom triangular and Delaunay networks, close to the isostatic point the parameter $\ell_0^\ast$ exhibits a linear dependence on $\Delta z$ with a negative coefficient (\autoref{fig:z-scaling}a,b), which has also been observed in 2D packing-derived networks \cite{Merkel2019}.

We also examined the $z$-scaling exponents of $a_{\ell}$ and $b$ close to isostaticity (\autoref{fig:z-scaling}d,e,g,h). 
For $a_\ell$, we find for phantom triangular networks a scaling exponent of $\approx0.2$, for Delaunay networks, we find an exponent close to $-1$, while we found an exponent of $-0.5$ in earlier work for packing-derived networks.
Meanwhile for $b$, we find for phantom triangular networks an exponent of $\approx -0.5$ or smaller, for Delaunay networks an exponent of roughly $-2$, while we have found before for packing-derived networks an exponent of $-1$.
Hence, the scaling exponent of both parameters strongly depend on network class.

Note that for a small fraction of the Delaunay networks, we did not observe a linear scaling between $\sigma_\ell$ and $\bar\ell-\ell_0$, suggesting that the linear relation between $\bar\ell_\mathrm{min}$ and $\sigma_\ell$ might be violated for these networks (appendix~\ref{sec:parameter extraction}).
A more detailed examination suggests that this could quite possibly be due to finite numerical cutoffs required to identify the transition point, which would make us miss the regime where this scaling is linear (appendix~\ref{sec:Delaunay issue}).
We excluded these networks from the averages shown in \autoref{fig:z-scaling}.
We stress that we only found deviations from the linear $\bar\ell_\mathrm{min}$ scaling for Delaunay networks with harmonic springs, while we could numerically confirm the predicted linear scaling for all phantom triangular and Voronoi networks, as well as the honeycomb network.

\subsection{The shear modulus scales linearly with isotropic stress.}\label{sec:shear modulus scales linearly}
In the previous sections (\ref{sec:mechanic properties} and \ref{sec:floppy-rigid transition and linear relation}), we showed that the shear modulus $G$ scales linearly with the isotropic strain beyond the transition point, $\Delta\varepsilon=\varepsilon-\varepsilon^\ast$. Moreover, a finite bulk modulus discontinuity at $\varepsilon^\ast$ implies that the isotropic stress $T$ also scales linearly with $\Delta\varepsilon$ to lowest order, which can be derived form Eqs.~\eqref{eq:l0 epsilon} and \eqref{eq:P prediction}. Hence, we would expect from the analytical predictions in \autoref{sec:analytics} that the shear modulus scales linearly with the isotropic stress:
\begin{equation}
G \sim T^{\alpha}\qquad\text{with $\alpha=1$.}
\end{equation}
However, recent numerical work has suggested different values for $\alpha$. For instance,  reference~\cite{Arzash2019} studied networks with rope-like potentials, and for $z=3.2$ the results suggested an exponent of $\alpha\approx0.85$ for phantom triangular and $\alpha\approx0.9$ for Delaunay networks, while $\alpha=1$ was found for honeycomb and Voronoi networks.

To resolve this contradiction between the numerical results from Ref.~\cite{Arzash2019} and our analytical results from \autoref{sec:analytics} and Ref.~\cite{Merkel2019}, we simulate here different kinds of rope-like networks with a high numerical precision, where we vary linear system size $W$ by more than an order of magnitude.
\autoref{fig:G_P}a shows the scaling of the shear modulus $G$ against the isotropic stress $T$, both averaged over 50 realizations, for phantom triangular networks, where we used two protocols. The open symbols correspond to a protocol without any shear stabilization. This means that no shear strain was applied after the creation of the network, and the $\varepsilon$ sweep was carried out at $\hat\gamma=0$. 
The closed symbols correspond to a protocol where we used shear stabilization when searching for the transition point, and as a consequence the $\varepsilon$ sweep was carried out at $\gamma=0$ (see \autoref{sec:effect of shear strain}).

We find that indeed, for the protocol with shear stabilization (closed symbols), the shear modulus $G$ scales linearly with isotropic stress $T$ over many orders of magnitude (\autoref{fig:G_P}a for phantom triangular networks \& inset for the other network classes). This observation is independent of system size.
However, without shear stabilization (open symbols), at small stress $T$ we observe a plateau, whose value depends on system size. Away from the plateau regions the curves largely collapse for different system sizes.

The appearance of a plateau in $G(T)$ in the protocol without shear stabilization can be readily understood from our analytical results.  Eq.~\eqref{eq:G prediction} states that $G$ is proportional to $\ell_0^\ast-\ell_0+b\gamma^2$, where $(\ell_0^\ast-\ell_0)\sim\Delta\varepsilon\sim T$ and $\gamma=\hat\gamma-\Delta\gamma_0$. Without shear stabilization, $\hat\gamma=0$ and so $\gamma=-\Delta\gamma_0$. This implies a plateau in $G$ that is proportional to $\Delta\gamma_0^2$. In other words, the plateau in $G(T)$ is related to the small anisotropy in the disordered networks. Shear stabilization removes this anisotropy and thus also the plateau in $G(T)$.

To test whether the plateaus that we find in \autoref{fig:G_P}a do not result form an averaging effect, we plot in \autoref{fig:G_P}b the same curves for individual realizations for a given system size.  We find that the plateau is also present in individual simulations, and that its height fluctuates across realizations. This makes sense, because the network anisotropy $\Delta\gamma_0$ also fluctuates across realizations.  Moreover, we find that the variance of $\Delta\gamma_0$ decreases inversely proportional to the number of springs in the system (\autoref{fig:G_P}b), which scales as $\sim W^2$. 
Hence, the plateau in $G(T)$ corresponds to a finite-size effect. A similar conclusion was drawn also in Ref.~\cite{Damavandi2021b} following a different line of argument. 

\begin{figure}  
\centering	\includegraphics[width=1\columnwidth]{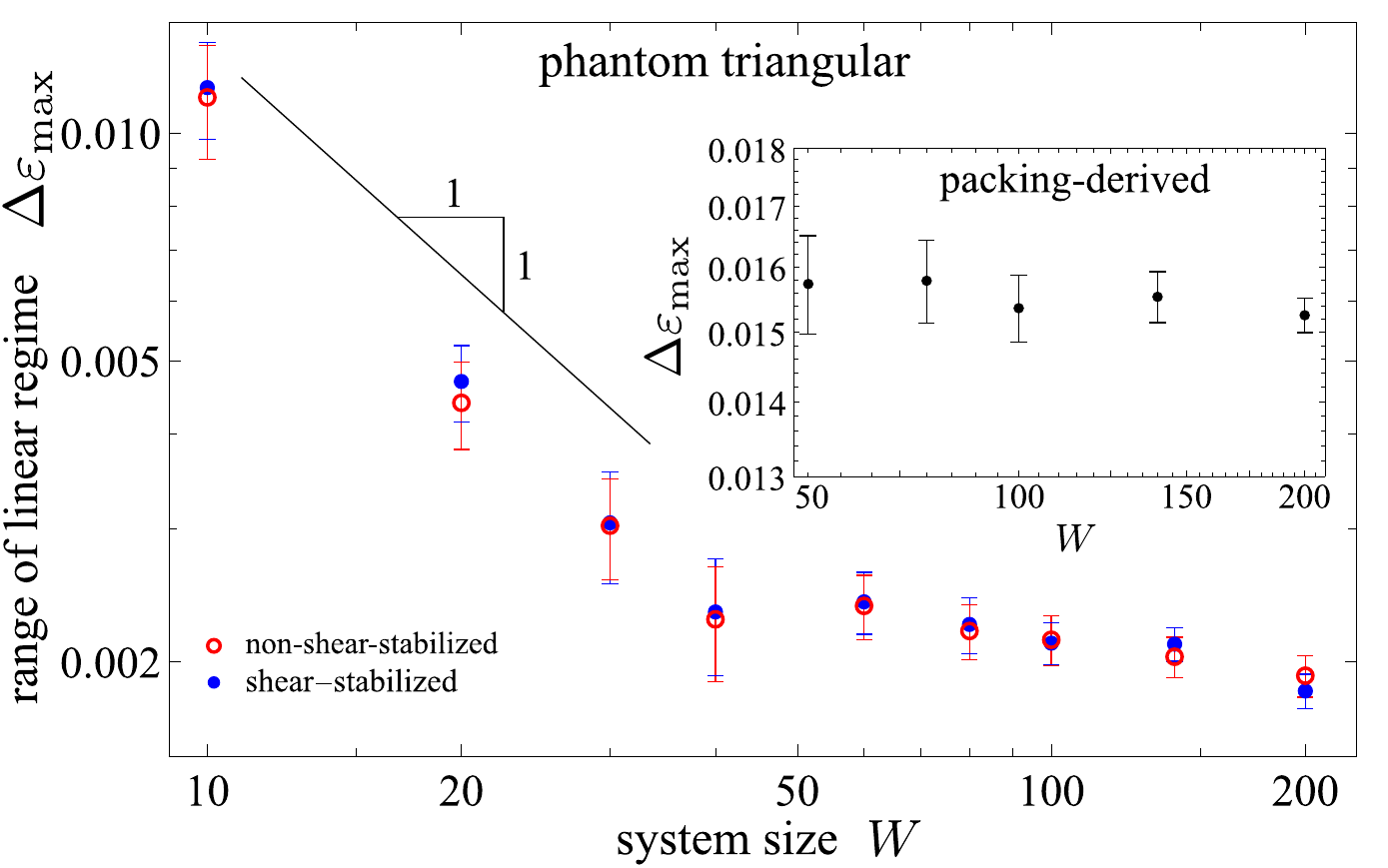}
	\caption{Plot of the isotropic strain range $\Delta\varepsilon_{\mathrm{max}}=\varepsilon_{\mathrm{max}}-\varepsilon^\ast$ within which the $\bar\ell_\mathrm{min}$ function scales linearly with $\sigma_\ell$, shown here for increasing system size $W$ for phantom triangular networks, and for shear-stabilized packing-derived networks (inset). In both cases, we use $z=3.2$ and rope-like potentials.
	The range $\Delta\varepsilon_{\mathrm{max}}$ is quantified as explained in appendix~\ref{sec:finite-size effects} and \autoref{fig:defEpsimax}.
	While for $W\gtrsim60$, the linear range of the phantom triangular networks appear to show a weak power-law dependence on $W$, there is no significant dependence on system size for packing-derived networks.
	Error bars indicate the standard error of the mean.
}\label{fig:upper bound for linear relation}
\end{figure}

\subsection{Linear range of $\bar\ell_\mathrm{min}$ shows no or weak system-size dependence}\label{sec:not from finite-size effect}
In recent work, it was pointed out that scaling exponents in spring networks under shear strain may be affected by finite-size effects \cite{Arzash2020}.  In particular, it was suggested that for networks of size $W$, finite-size effects could affect scaling exponents when shearing the systems by $\Delta\gamma\lesssim W^{-1/\nu}$ beyond the transition point $\gamma^\ast$, where $\nu>0$. This would correspond to a diverging length scale $\xi\sim\vert\Delta\gamma\vert^{-\nu}$.
While in this article, we probe the system mechanics with respect to isotropic strain $\varepsilon$ instead, we still wanted to check whether such finite-size effects could affect our results.

The system mechanics with respect to isotropic strain is crucially determined by how the minimal-length function $\bar\ell_\mathrm{min}$ scales with $\sigma_\ell$ in Eq.~\eqref{eq:Lmin expansion} (\autoref{sec:analytics}, appendix \ref{sec:lmin general}).
Hence, we were wondering whether the linear scaling of $\bar\ell_\mathrm{min}$ with $\sigma_\ell$ is only valid close to the transition point with strains $\Delta\varepsilon<\Delta\varepsilon_\mathrm{max}\sim W^{-1/\nu_\ell}$ for some $\nu_\ell>0$. In other words, we wondered whether the range $\Delta\varepsilon_\mathrm{max}$ of linear $\bar\ell_\mathrm{min}$ scaling would algebraically decrease to zero with increasing system size $W$.

In \autoref{fig:upper bound for linear relation} we show the resulting dependency of $\Delta\varepsilon_\mathrm{max}$ on system size $W$ for both phantom triangular networks and packing-derived networks, both with rope-like spring potentials, where $\Delta\varepsilon_\mathrm{max}$ is quantified as described in appendix~\ref{sec:finite-size effects}.
For the phantom networks, beyond an initial quick decrease in $\Delta\varepsilon_\mathrm{max}$ for small $W$, we find that for $W\geq60$ our data indicates a finite-size scaling exponent in the range $1/\nu_\ell\in [0.04, 0.35]$. Thus, the range of linear scaling in the $\bar\ell_\mathrm{min}$ function slowly decreases with system size.
In contrast, for the packing-derived networks discussed in  Ref.~\cite{Merkel2019}, we find a range of $1/\nu_\ell\in [-0.03, 0.08]$. This means that the linear scaling range of $\bar\ell_\mathrm{min}$ is subject to none, or at most a weak finite-size effect.
Thus, intriguingly, the effect of finite-size effects on the linear scaling range of $\bar\ell_\mathrm{min}$ appears to depend on the class of network studied.

Given that strain-controlled transitions in spring networks have been shown to be critical transitions \cite{Feng2016,Sharma2016a,Sharma2016b,Shivers2019,Arzash2019,Arzash2020}, we wondered whether we would also observe a divergence in the fluctuations close to the transition.  
Focusing on the scaling behavior of a non-affine motion parameter $\Gamma$ with system size $W$ and distance to the transition point, $\varepsilon-\varepsilon^\ast$, we find indeed such a divergence (appendix~\ref{sec:finite-size effects}).
Moreover, we also find finite-size effects with exponents of $1/\nu_\Gamma\approx 0.75$ for phantom triangular networks and $1/\nu_\Gamma\approx 0.6$ for packing-derived networks.
Intriguingly, these exponents are very different from what we observe for the linear scaling regime of $\bar\ell_\mathrm{min}$, i.e.\ $\nu_\Gamma\neq\nu_\ell$. This suggests that the linear range of $\bar\ell_\mathrm{min}$ is not controlled by the diverging length scale that controls non-affine motion.
A possible reason for this is that close to the transition the non-affinity parameter $\Gamma$ mostly captures motions that are (to first-order) unconstrained by spring lengths, while $\bar\ell_\mathrm{min}(\sigma_\ell)$ characterizes spring length behavior.

\section{Discussion}
We studied the elastic behavior of sub-isostatic spring networks that are rigidified by isotropic expansion, comparing numerical simulation results with analytical predictions from Ref.~\cite{Merkel2019}. 
We first summarized the approach from Ref.~\cite{Merkel2019}, which proposed an analytical framework to predict the elastic network properties using a minimal-length function $\bar\ell_\mathrm{min}$  (Eq.~\eqref{eq:lmin with shear}).
This minimal-length function allows to map the physical problem of the strain-induced stiffening transition to the purely geometric problem of finding a minimal length.
We show that the way $\bar\ell_\mathrm{min}$ changes with length fluctuations and shear strain directly defines the mechanical properties of under-constrained spring networks.
Indeed, close to the transition, $\bar\ell_\mathrm{min}$ is predicted to scale linearly with spring length fluctuations, in a way that is directly linked to the SSS that is created at the transition \cite{Merkel2019}.

The $\bar\ell_\mathrm{min}$ formalism allows to make several quantitative predictions of the elastic network behavior close to the transition \cite{Merkel2019}.
These predictions include the coefficient describing the shape of the rigid-floppy boundary with respect to shear and isotropic strain, the value of the bulk modulus discontinuity at the transition, the linear scaling coefficient of shear modulus with isotropic tension, the value of the shear modulus discontinuity for networks under shear strain, the coefficient of the linear shear modulus scaling beyond this transition, and the coefficient describing the anomalous Poynting effect.
Because all these predictions are based only on the three parameters $\ell_0^\ast$, $a_{\ell}$ and $b$, by combining these predictions one can construct non-trivial parameter-free predictions that apply to any athermal under-constrained material \cite{Merkel2019}.

Here, we numerically verified the predicted linear scaling of the minimal-length function near the transition and extracted the three parameters $\ell_0^\ast$, $a_{\ell}$ and $b$ for four different network classes, including phantom triangular, Delaunay, honeycomb, and Voronoi networks. 
Based on these parameters, we compute the bulk modulus discontinuity $\Delta B$, which predicts well our numerical results for all network classes (\autoref{fig:floppy-rigid transition}). Moreover, we also recovered the predicted linear scaling of the shear modulus $G$ with isotropic tension $T$ close to the transition. 

Next we explored the scaling of the parameters $\ell_0^\ast$, $a_{\ell}$ and $b$ with respect to connectivity $z$. 
We found that the scaling of the parameters $a_{\ell}$ and $b$ with the distance to isostaticity $\Delta z=4-z$ strongly depends on the network class. The scaling exponent for $a_{\ell}$ can even change sign, varying from $-1$ for Delaunay networks to $\approx 0.2$ for phantom triangular networks (\autoref{fig:z-scaling}, with an exponent of $-0.5$ for packing-derived networks \cite{Merkel2019}).
The scaling exponent for $b$ varies from $-2$ for Delaunay networks to $-0.5$ for phantom triangular networks (\autoref{fig:z-scaling}, with an exponent of $-1$ for packing-derived networks \cite{Merkel2019}).
This dependency on network class is not too surprising, since the parameters $\ell_0^\ast$, $a_{\ell}$ and $b$ depend on the microscopic network structure, which varies with network class.
In contrast, the value of $\ell_0^\ast$ always showed a linear dependency on $\Delta z$, where intercept and slope depend on network class.

One prediction of the formalism in Ref.~\cite{Merkel2019} is a linear scaling of the shear modulus $G$ with the isotropic stress $T$ close to the transition point: $G \sim T^{\alpha}$ with $\alpha=1$.
This is a direct consequence of the discontinuity $\Delta B$ in the bulk modulus and of the linear scaling of the shear modulus $G$ with strain $\Delta\varepsilon$.
Note that the linear scaling of $G$ with $\Delta\varepsilon$ is essentially a consequence of the linear scaling of the $\bar\ell_\mathrm{min}$ function with $\sigma_\ell$.
While in \autoref{sec:effect of shear strain} we make the assumption that $\bar{\ell}_\mathrm{min}$ at its minimum is analytical in $\gamma$, non-analytic behavior in $\gamma$ would still lead to an integer exponent $\alpha$ (appendix~\ref{sec:lmin general}).
However, the prediction of integer $\alpha$ seems to be at odds with recent numerical work, which suggested that the value of $\alpha$ can be different from one for networks with a rope-like interaction potential, depending on the disordered nature of the network \cite{Arzash2019}.
In particular, Ref.~\cite{Arzash2019} found an exponent of $\alpha\approx0.85$ for phantom triangular networks and $\alpha\approx0.9$ for Delaunay networks with connectivity of $z=3.2$.

To reconcile the two results from Refs.~\cite{Arzash2019,Merkel2019}, we numerically studied the $G(T)$ scaling with an increased numerical precision, and our results confirmed the analytically predicted scaling exponent of $\alpha=1$ in both phantom triangular and Delaunay networks (\autoref{fig:G_P}a and inset).
We show that the result also depends on a small random anisotropy in the generated network. 
In the presence of such a finite anisotropy, we observed a plateau in the shear modulus $G(T)$ for small isotropic stress $T$, consistent with the analytic prediction, Eq.~\eqref{eq:G prediction}.
This plateau disappears when using shear stabilization \cite{Dagois-Bohy2012}, which removes the network anisotropy by shearing the network by a shear strain $\Delta\gamma_0$ (\autoref{sec:effect of shear strain}).
We moreover show that the plateau disappears for larger system sizes (\autoref{fig:G_P}b inset).
Hence, while without shear stabilization large system sizes are required to probe the behavior close to the transition, shear stabilization allows to explore this regime already for smaller systems.

We see two possible reasons for the discrepancy in the $G(T)$ scaling between Refs.~\cite{Arzash2019,Merkel2019}.
First, we used the conjugate gradient minimizer code developed in Ref.~\cite{Merkel2019}, which allows us to probe the system at least two orders of magnitude closer to the transition point than Ref.~\cite{Arzash2019} (see gray vertical line in \autoref{fig:G_P}a). For instance in phantom triangular networks we observe an exponent of $\alpha<1$ for larger isotropic stress $T\gtrsim10^{-3}$, which seems consistent with the value of $0.85$ given by Ref.~\cite{Arzash2019}, while we observe an exponent of $\alpha=1$ for stress $T$ smaller than that.
Second, we show that a small anisotropy in the generated network can lead to a plateau in the shear modulus curve $G(T)$, which could in turn affect the inferred scaling exponent. 

Previous work suggested that finite-size effects could affect scaling exponents in spring networks \cite{Arzash2020}.
This can occur whenever the system size is on the order of or smaller than a length scale that diverges close to the transition point.
While Ref.~\cite{Arzash2020} focused on shear simulations, we wanted to test whether such an effect could also arise in our isotropic-strain simulations.
To this end, we numerically tested in what range around the transition point the linear scaling of the $\bar\ell_\mathrm{min}$ function holds.
Our results suggest that this potentially depends on the class of network studied. While in phantom triangular networks, 
this range decreases weakly with system size, we did not find a significant decrease in packing-derived networks.
This is also consistent with \autoref{fig:G_P}a, which suggests that $G(T)$ is largely independent of system size $W$ for the range of $W$ probed.

Intriguingly, we found that non-affine motion $\Gamma$ shows a much stronger system-size dependence, suggesting that it is controlled by a length scale that does not affect the linear range of $\bar\ell_\mathrm{min}$. One possible reason for this is that close to the transition point non-affine motions are to linear order unconstrained in under-constrained networks. 
Better understanding this difference in the finite-size scaling behavior of $\Gamma$ and $\bar\ell_\mathrm{min}$ is an interesting avenue for future research.

\section*{Acknowledgements}
We thank Martin Lenz for fruitful discussions.
We thank the Centre Interdisciplinaire de Nanoscience de Marseille (CINaM) for providing office space.
The project leading to this publication has received funding from the ``Investissements d'Avenir'' French Government program managed by the French National Research Agency (ANR-16-CONV-0001) and from ``Excellence Initiative of Aix-Marseille University - A*MIDEX''.
The Centre de Calcul Intensif d’Aix-Marseille is acknowledged for granting access to its high-performance computing resources.

\begin{appendix}

\section{Generalization to networks with heterogeneous spring constants and rest lengths}\label{sec:heterogeneous}
In the main text, we focused on the case where all springs share the same rest length $\ell_0$ and the same spring constant $k$. Here, we generalize this to networks where rest length and spring constant may differ among the springs, as in Eq.~\eqref{eq:E_general}. 
Similar to Ref.~\cite{Merkel2019}, we introduce re-scaled spring lengths $\tilde{\ell}_i$, re-scaled spring constants $\tilde{k}_i$, and an average spring rest length $\ell_0$ in a way that allows us to rewrite Eq.~\eqref{eq:E_general} in the form:
\begin{equation}
    e = \sum_{i=1}^N{\tilde{k}_i\big(\tilde{\ell}_i-\ell_0\big)^2}.\label{eq:rescaled spring lengths energy}
\end{equation}
For this to work, we need to define the re-scaled spring lengths as
\begin{equation}\label{eq:rescale L factor}
\Tilde{\ell}_i=  \ell_i\frac{\ell_0}{\ell_{0i}}.
\end{equation}
This will accordingly give rise to a new re-scaled spring constants
\begin{equation}\label{eq:rescaled K}
\Tilde{k}_i=k_i\left( \frac{\ell_{0i}}{\ell_0} \right)^2 .
\end{equation}
Finally, we choose to define $\ell_0$ as the quadratic mean of the $\ell_{0i}$, weighted by the $k_i$:
\begin{equation}\label{eq:L0 in hetero case}
\ell_0 = \sqrt{\frac{\sum_i k_i \ell_{0i}^2}{\sum_i k_i}}.
\end{equation}
Using the re-scaling Eqs.~\eqref{eq:rescale L factor}--\eqref{eq:L0 in hetero case}, Eq.~\eqref{eq:E_general} can be exactly re-expressed as Eq.~\eqref{eq:rescaled spring lengths energy}.
This network energy can be transformed into
\begin{equation}\label{eq:E two terms hetero}
e = N \left[ \left( \bar{\ell}- \ell_0 \right)^2 + \sigma_{\ell}^2 \right] .
\end{equation}
Here, $\bar{\ell}$ and $ \sigma_\ell$ are defined as the average and standard deviation of the re-scaled spring length $\tilde{\ell}_i$ with the weighting factors $\tilde{k}_i$:
\begin{align}
\bar{\ell}:= \frac{\sum_i \tilde{k}_i\tilde{\ell}_i}{\sum_i \tilde{k}_i} ;\ 
\sigma_{\ell}^2:= \frac{\sum_i \tilde{k}_i\left(\bar{\ell}-\tilde{\ell}_i\right)^2}{\sum_i \tilde{k}_i}.
\end{align}
The subsequent discussion in sections~\ref{sec:key_idea}--\ref{sec:mechanic properties} remains unchanged. 

\section{Network generation}
\label{sec:network generation}
Networks were created using the following protocols.

\textit{Phantom triangular}  (\autoref{fig:floppy-rigid transition}a) \cite{Broedersz2011b}: 
Following Ref.~\cite{Arzash2019}, a 2D triangular lattice of spacing 1 is first constructed by depositing three sets of $W$ parallel filaments each at angles of $0^{\circ}$, $60^{\circ}$ and $120^{\circ}$ with the $x$-axis, respectively. 
To reduce the connectivity from $z=6$ to values observed in e.g.\ collagen networks \cite{Lindstrom2010} of $3\leq z<4$, we first detach at each node one filament, which is randomly chosen among the three crossing filaments. This creates a network of homogeneous connectivity $z=4$. To avoid system-spanning filaments, one spring is removed at a random position on each filament, giving the average connectivity $z=4-6/W$.
To further reduce the connectivity to a defined value $z$, we implement an iterative procedure. At each iteration, we randomly remove only a few of the springs and then clear off all of the dangling springs and isolated islands. This is repeated until the desired connectivity $z$ is reached.

\textit{Delaunay} (\autoref{fig:floppy-rigid transition}b):
Delaunay networks are constructed from $W^2$ nodes that are placed at uncorrelated random positions in a square box of side $W$. The connectivity of initially $z=6$ is decreased to the desired value $z$ by employing the same protocol using random cuts as for the phantom triangular networks.

\textit{Honeycomb} (\autoref{fig:floppy-rigid transition}c):
We construct a network of $W^2/3$ regular hexagons with side length 1.

\textit{Voronoi} (\autoref{fig:floppy-rigid transition}c): 
Voronoi networks correspond to the Voronoi tessellation of $W^2/2$ nodes at uncorrelated random positions in a square box of side $W$. 

In all four network classes, we set the dimensionless spring rest lengths $\ell_{0i}$ to the respective initial spring lengths before any deformation is applied, i.e.\ at $(\varepsilon, \hat\gamma)=(0,0)$. We set the dimensionless spring constants as the inverse of the respective rest length at zero strain, $k_i=1/\ell_{0i}$.

\section{Details of numerical strain sweeps and computation of the elastic moduli}
\label{sec:details sweeps}

In this paper, we exclusively carry out sweeps of isotropic strain $\varepsilon$. Before each sweep, we first identified the transition point $\varepsilon^\ast$. 
To this end, we implemented a bisection scheme, which we optimized by linearly interpolating the transition point in each step.
We defined networks as rigid whenever their isotropic stress $T$ is above a cutoff value of $10^{-10}$ (two orders of magnitude above the tolerance for the residual force cutoff per degree of freedom, $10^{-12}$). We use isotropic stress $T$ as a criterion for network rigidity, because it is much faster to compute than an elastic modulus.

In the bisection to identify the transition point, we also implemented the option to perform shear stabilization to remove network anisotropy (sections~\ref{sec:effect of shear strain} and \ref{sec:network generation and minimization}). This is done by treating the shear strain $\hat{\gamma}$ as an additional degree of freedom during each energy minimization of the bisection process. In any case, shear stabilization was always turned off (i.e.\ shear remains constant) after the transition point $\varepsilon^\ast$ has been identified.

We apply an exponential sweep of isotropic strain to probe the scaling behavior of network mechanics close to the transition point $\varepsilon^\ast$. 
In particular, we probed strain values $\varepsilon-\varepsilon^\ast=10^{-10 + 0.2 k}$, where the step index $k$ ranged from 0 to 51 by default, with only two exceptions. First, in \autoref{fig:floppy-rigid transition}, we apply the same exponential sweep, yet with $k$ ranging from 0 to 7 only, which is then followed by a linear sweep.
Second, for the Voronoi networks of size $W=70$ (\autoref{fig:z-scaling} and \autoref{fig:G_P}a inset) we needed to increase the residual force cutoff for the energy minimization to $10^{-10}$, and so we also raised the cutoff in isotropic stress $T$ to identify the transition point to $10^{-8}$. Accordingly, we changed the sweep to the values $\varepsilon-\varepsilon^\ast=10^{-8+0.2 k}$ with $k$ ranging from 0 to 41. 

We computed the elastic moduli using two different methods. 
For not too big networks, we diagonalized the Hessian of the system energy and used the resultant eigenvalues to compute elastic moduli  \cite{Huang1950,Born1955,Lemaitre2006,Merkel2018,Merkel2019}. This approach produces a higher numerical precision and was suitable for typical system sizes $W<100$ (\autoref{fig:floppy-rigid transition} and \ref{fig:z-scaling}).
However, in \autoref{fig:G_P} we studied networks with a large system size, and so we used a less time-intensive way of computing the shear modulus $G$.
We computed $G$ through a difference quotient of the shear stress over the shear strain: $G(\varepsilon, \gamma=0)=[\sigma(\varepsilon, \Delta)-\sigma(\varepsilon, -\Delta)]/2\Delta$, where we numerically tuned and found the optimized shear strain $\Delta=5\times10^{-5}$. We also noticed that for $\varepsilon-\varepsilon^\ast<10^{-7}$ the shear modulus computed with this method could deviate significantly from the true value. We hence excluded these data points in \autoref{fig:G_P} and the lowest isotropic stress there is accordingly $T \approx 10^{-7}$.  

\section{Extraction of the parameters $a_\ell$ and $b$ of the minimal-length function}
\label{sec:parameter extraction}

To extract $a_\ell$ from numerical data, one could just directly use the $\bar\ell_\mathrm{min}(\sigma_\ell)$ function (Eq.\eqref{eq:Lmin expansion}).
However, this approach depends on the correct identification of the transition point $\ell_0^\ast$. While we can identify $\ell_0^\ast$ with a relatively high precision of $\sim 10^{-10}$, we could even remove the dependency on $\ell_0^\ast$ entirely when determining $a_\ell$.
To this end, we note that in an energy-minimized state, the energy is also minimal with respect to variation of $\sigma_\ell$, i.e.\ $d e/d\sigma_\ell =0$. From Eq.~\eqref{eq:E two terms}, and using the insight that $\bar\ell=\bar\ell_\mathrm{min}(\sigma_\ell)$ in the rigid regime, the minimization condition reads:
\begin{align}
\frac{d e}{d \sigma_\ell}
&=2 N \left[ \left(\bar\ell-\ell_0\right)\frac{d\bar\ell_\mathrm{min}}{d\sigma_\ell} + \sigma_\ell \right] =0.
\end{align}
Using Eq.~\eqref{eq:Lmin expansion}, the derivative of the minimal-length function is $d\bar\ell_\mathrm{min}/d\sigma_\ell=-a_\ell$. Taken together, we thus obtain the linear relation:
\begin{equation}\label{eq:linear equivalence}
\sigma_\ell=a_\ell\left( \bar\ell-\ell_0 \right) .
\end{equation}
Based on Eq.~\eqref{eq:linear equivalence}, examining the relation between $\sigma_\ell$ and $(\bar\ell-\ell_0)$ (e.g.\ \autoref{fig:z-scaling}d inset) allows both to effectively verify the scaling of the minimal-length function (Eq.~\eqref{eq:Lmin expansion}), and to extract the value of $a_\ell$.  This approach does not involve the critical value $\ell_0^\ast$ which we obtain with a lower precision as compared to $\bar\ell$ and $\sigma_\ell$ (as precise as $10^{-12}$).

To extract the parameter $b$, we use the derived shear modulus formula (Eq.~\eqref{eq:G prediction}), instead of directly using the original minimal-length function (Eq.~\eqref{eq:lmin with shear}) since we do not shear the networks (i.e.\ $\gamma=0$). As before, we intend not to use the critical value $\ell_0^\ast$. Thus, to replace the term $(\ell_0^\ast-\ell_0)$ that appears in the shear modulus formula (Eq.~\eqref{eq:G prediction}), we insert Eq.~\eqref{eq:linear equivalence} back into the minimal-length function (Eq.~\eqref{eq:Lmin expansion}) and obtain
\begin{equation}
\ell_0^\ast-\ell_0=\left(1+a_\ell^2\right)\left(\bar\ell-\ell_0\right) .
\end{equation}
Combining this equation with the shear modulus formula (Eq.~\eqref{eq:G prediction}) yields
\begin{equation}\label{eq:b fit}
G=4b\left(\bar\ell-\ell_0\right).
\end{equation}
We used this equation to extract $b$ from the plots of $G$ over $\bar\ell-\ell_0$ (\autoref{fig:z-scaling}g inset).

By default we use the first 25 data points from an exponential sweep to numerically fit Eqs.~\eqref{eq:linear equivalence} and \eqref{eq:b fit} and extract the parameters $a_\ell$ and $b$.
Note that $a_\ell$ in \autoref{fig:upper bound for linear relation} is defined in the very same way, based on the first 25 data points of an exponential sweep.
Meanwhile in \autoref{fig:floppy-rigid transition} we use only the first 5 data points due to a decreased step number $n$ in the exponential sweep of isotropic strain (appendix~\ref{sec:details sweeps}).

We note that rope-like spring potentials can be treated as well with the analytical framework in \autoref{sec:analytics}, which we took into account when computing $\bar{\ell}$ and $\sigma_{\ell}$ here.
While one way to treat rope-like spring potentials was discussed in Ref.~\cite{Merkel2019}, where each spring is subdivided into a series of shorter springs, we chose here an alternative approach.
We used the fact that for rope-like spring potentials, a spring $i$ only affects the mechanics when $\ell_{i}>\ell_{0i}$, while springs with $\ell_{i}<\ell_{0i}$ do not contribute.  Thus, to compute $\bar{\ell}$ and $\sigma_{\ell}$, whenever for any spring $i$ the distance of the two connected nodes is smaller than the rest length $\ell_{0i}$, we set the spring length to $\ell_{i}=\ell_{0i}$. 
This redefinition of $\ell_i$ does not affect the computation of shear modulus $G$ and tension $T$.

\section{Any Voronoi network at creation has a state of self stress.}
\label{sec:Voronoi self-stress}

\begin{figure}[t]  
\centering	\includegraphics[width=0.6\columnwidth]{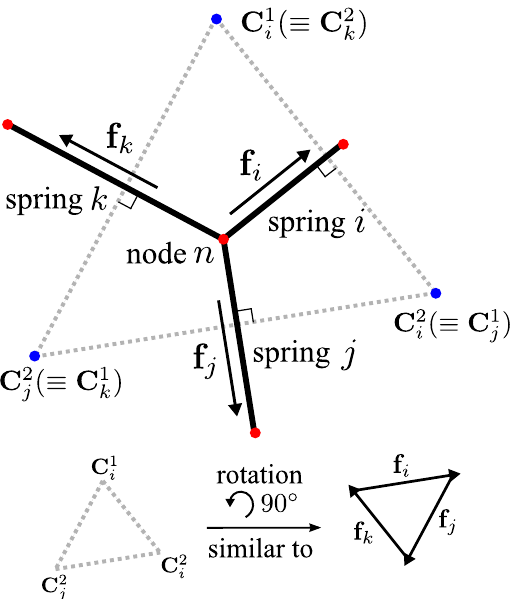}
	\caption{Illustration of local force balance at any node $n$ of a Voronoi network, demonstrating that Voronoi networks have a SSS at creation. Red dots are the internal nodes, while blue dots are the neighboring random seeds used for the Voronoi tessellation. Geometrically, node $n$ is created as the circumcenter of the local triangle (gray dashed lines) formed by these seeds, and the three local springs $i$, $j$ and $k$ (black segments) are the vertical bisectors of the respective sides. The vectors $\bm{C}_i^1$ and $\bm{C}_i^2$ refer to the two seed points at the side perpendicular to spring $i$ (and similarly for springs $j$ and $k$). The node $n$ is force-balanced when the magnitude of the spring tensile forces ($\bm{f}_i$, $\bm{f}_j$ and $\bm{f}_k$, in black arrows) follow the proportionality relation, Eq.~\eqref{eq:tension proportionality}. These forces will form a closed triangle that is similar to the local triangle by a rotation of $90^\circ$, thus giving zero net force, see also Eqs.~\eqref{eq:tensile force balance} and \eqref{eq:local force balance}.
	}
	\label{fig:Voronoi_selfStress}
\end{figure}

We numerically found that Voronoi networks have a critical isotropic strain very close to zero, $\varepsilon^\ast\approx 0$. Here we show that the critical strain is indeed exactly zero, by proving that there is a state of self stress right at creation of these networks. In other words, at creation ($\varepsilon=0$) these networks can sustain finite tensions in a subset of springs, while force balance is maintained at the internal nodes. 

The geometric structure of a Voronoi network allows for the following set of spring tensions $t_i$ (with $i$ being a spring index) to be a state of self stress:
\begin{equation}\label{eq:tension proportionality}
    t_i = \eta \vert\bm{C}_i^2 - \bm{C}_i^1\vert.
\end{equation}
Here, $\eta$ is some constant factor, the vectors $\bm{C}_i^{1}$ and $\bm{C}_i^2$ refer to the two Voronoi seed points that are closest to spring $i$ (\autoref{fig:Voronoi_selfStress}; i.e.\ $\bm{C}_i^{1}$ and $\bm{C}_i^2$ are the two points that generated the line that defines spring $i$), and $\vert\cdot\vert$ denotes the length of a vector.

\begin{figure*}
    \centering	
    \includegraphics[width=1.8\columnwidth]{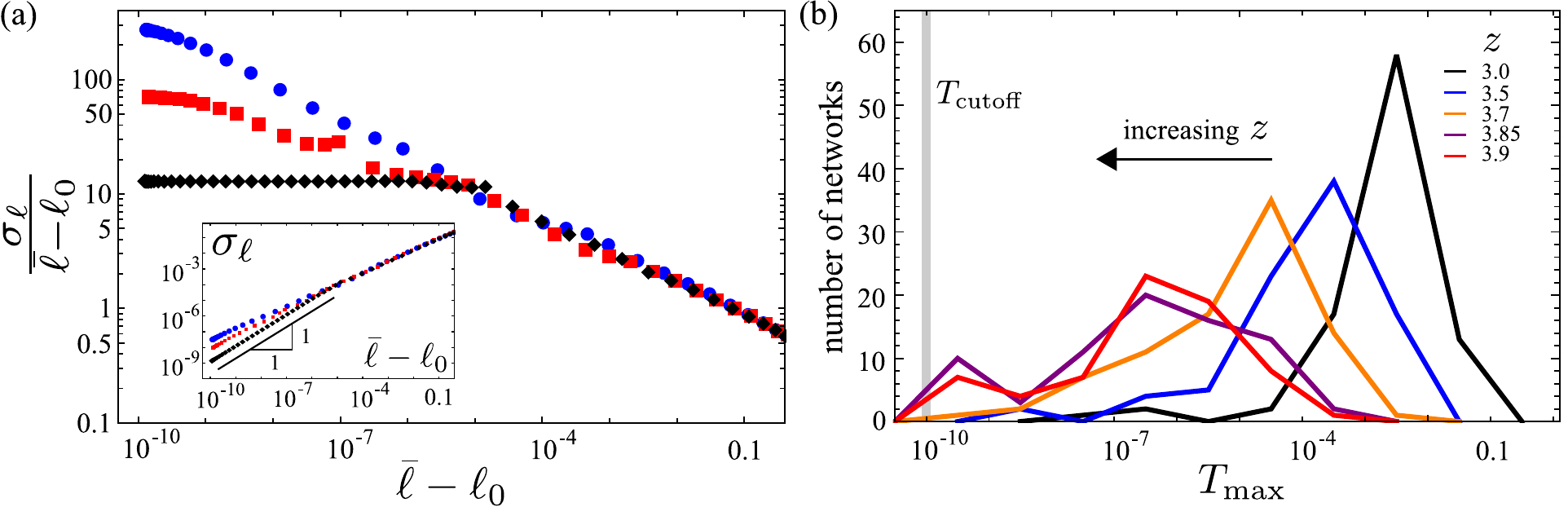}
	\caption{(a) The quotient $\sigma_\ell/(\bar\ell-\ell_0)$ plotted versus $\bar\ell-\ell_0$ for Delaunay networks with harmonic spring potentials, shown here for three networks. These networks show either a linear (black diamonds) or non-linear (blue dots and red squares) scaling between $\sigma_\ell$ and $\bar\ell-\ell_0$ close to the transition point (inset), where a linear scaling is reflected by a plateau in the quotient $\sigma_\ell/(\bar\ell-\ell_0)$.
	(b) Histograms of the extent $T_\mathrm{max}$ of the plateau in the quotient $\sigma_\ell/(\bar\ell-\ell_0)$ for Delaunay networks with harmonic spring potentials of different connectivity $z$. We compute $T_\mathrm{max}$ as the isotropic stress value where the quotient $\sigma_\ell/(\bar\ell-\ell_0)$ starts to show a deviation of more than 10 \% from the one computed at the detected transition point. 
    }
	\label{fig:Delaunay non-linear}
\end{figure*}

To show that the set of spring tensions $t_i$ form a state of self stress, we demonstrate that they satisfy local force balance at each node. To this end, we focus here on a node $n$ that is connected to springs $i, j, k$ as shown in \autoref{fig:Voronoi_selfStress}. The force that spring $i$ exerts on the node $n$ is $\bm{f}_i=t_i\bm{e}_i$, where $\bm{e}_i$ is the unit tangent vector of spring $i$ pointing away from node $n$.
Furthermore, we have 
\begin{equation}\label{eq:tensile force balance}
    \begin{aligned}
        \bm{f}_i &= t_i\bm{e}_i \\
        &= \eta\vert\bm{C}_i^2 - \bm{C}_i^1\vert\bm{e}_i \\
        &= \eta\vert\bm{C}_i^2 - \bm{C}_i^1\vert\,\frac{\bm{R}(\pi/2)\cdot(\bm{C}_i^2 - \bm{C}_i^1)}{\vert\bm{C}_i^2 - \bm{C}_i^1\vert} \\
        &= \eta\bm{R}(\pi/2)\cdot(\bm{C}_i^2 - \bm{C}_i^1).
    \end{aligned}
\end{equation}
Here, in the second line, we inserted the spring tensions Eq.~\eqref{eq:tension proportionality}. In the third line, we used the fact that spring $i$ is perpendicular to the segment connected by the two seed points $\bm{C}_i^1$ and $\bm{C}_i^2$, while the operator $\bm{R}(\pi/2)$ performs a counter-clockwise rotation by an angle of $\pi/2$. 
An analogous equation to Eq.~\eqref{eq:tensile force balance} holds also for the forces by springs $j$ and $k$.  As a consequence, the sum of these three forces is zero:
\begin{equation}\label{eq:local force balance}
    \bm{f}_i + \bm{f}_j + \bm{f}_k = 0.
\end{equation}
In other words, force balance on node $n$ holds.
This proof is also illustrated at the bottom of \autoref{fig:Voronoi_selfStress}; up to the factor of proportionality $\eta$, the three forces $\bm{f}_i,\bm{f}_j,\bm{f}_k$ correspond to the three triangle sides rotated by $\pi/2$, which is why they add up to zero. Hence, Voronoi networks at creation have a state of self stress given by Eq.~\eqref{eq:tension proportionality}.

\section{Apparent non-linear scaling of $\bar{\ell}_{\mathrm{min}}$ in some Delaunay networks} 
\label{sec:Delaunay issue}
For Delaunay networks with harmonic spring potentials, we observed that a fraction of the networks did not seem to follow the linear relation \eqref{eq:linear equivalence} between $\sigma_\ell$ and $(\bar\ell-\ell_0)$ (blue and red data points in \autoref{fig:Delaunay non-linear}a inset). 
This is also apparent from the absence of a plateau in $\sigma_\ell/(\bar\ell-\ell_0)$ (compare blue and red with black data points in \autoref{fig:Delaunay non-linear}a).
From our arguments in appendix~\ref{sec:parameter extraction}, it follows that this non-linearity also implies a non-linear scaling of the minimal-length function $\bar\ell_\mathrm{min}$ with $\sigma_\ell$, which would also affect the elastic network properties, Eqs.~\eqref{eq:P prediction}-\eqref{eq:G prediction}.

We wondered whether this non-linear scaling between  $\sigma_\ell$ and $(\bar\ell-\ell_0)$ was just due to finite numerical cutoffs, or whether it reflects the real scaling behavior infinitesimally close to the transition point. Numerical limitations arise because we cannot probe the networks arbitrarily close to the true transition point. Indeed, we used a cutoff value of $T_\mathrm{cutoff}=10^{-10}$ for the isotropic stress $T$ to numerically identify the transition point. In other words, at the detected transition point we are already in the rigid regime by some small extent beyond the true transition point.
If the plateau in $\sigma_\ell/(\bar\ell-\ell_0)$ exists only close to the true transition point until some isotropic stress value $T_\mathrm{max}<T_\mathrm{cutoff}$, we will not detect it since we missed that regime.  To test if this could be the case, we created histograms of the extent $T_\mathrm{max}$ of the plateau for different connectivity $z$ (\autoref{fig:Delaunay non-linear}b).  For a given network, we define $T_\mathrm{max}$ as the isotropic stress of the data point at which the quotient $\sigma_\ell/(\bar\ell-\ell_0)$ first deviates by more than 10\% from the value of this quotient at the detected transition point. For networks where the plateau ends below $T_\mathrm{cutoff}$, we would find with this approach $T_\mathrm{max}\approx T_\mathrm{cutoff}=10^{-10}$.
If there is a significant excess of networks where we numerically do not observe a plateau, this could be an indication that there is indeed no plateau. 

\autoref{fig:Delaunay non-linear}b shows that for Delaunay networks with harmonic springs, $T_\mathrm{max}$ generally decreases with connectivity $z$, and that we only observe a peak around $T_\mathrm{max}\approx T_\mathrm{cutoff}$ occur mostly for the two largest values of $z$. Even in these cases, the peak is not very pronounced and may very well arise from the integral of the real $T_\mathrm{max}$ distribution from 0 to $T_\mathrm{cutoff}$. In other words, these networks may possibly have a plateau which ends just too close to the transition point for us to detect it.

This is also consistent with the observation that most of these curves appear to collapse with the curves that do show a plateau beyond the end of the plateau (\autoref{fig:Delaunay non-linear}a). This suggests that the non-linear scaling regime just corresponds to a regime governed by higher-order terms.  In future work, it will be interesting to see if these higher-order terms could also be predicted from first principles.

\begin{figure}
    \centering	
    \includegraphics[width=0.9\columnwidth]{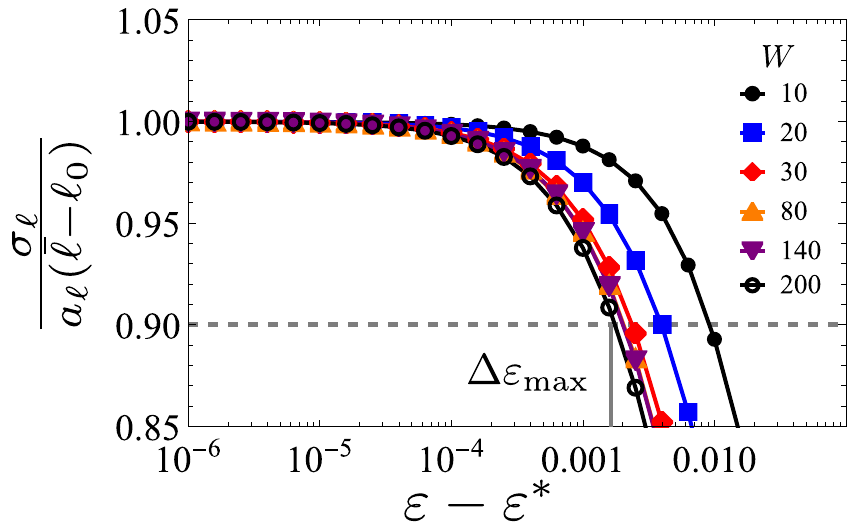}
	\caption{$\Delta\varepsilon_{\mathrm{max}}$ in Fig~\ref{fig:upper bound for linear relation} is defined as the strain range where the ratio $\sigma_{\ell}/(\bar{\ell}-\ell_0)$ shows a deviation of less than 10\% from its value $a_\ell$ defined very close to the transition point (appendix~\ref{sec:details sweeps}).
    }
	\label{fig:defEpsimax}
\end{figure}

\begin{figure*}
    \centering	
    \includegraphics[width=2\columnwidth]{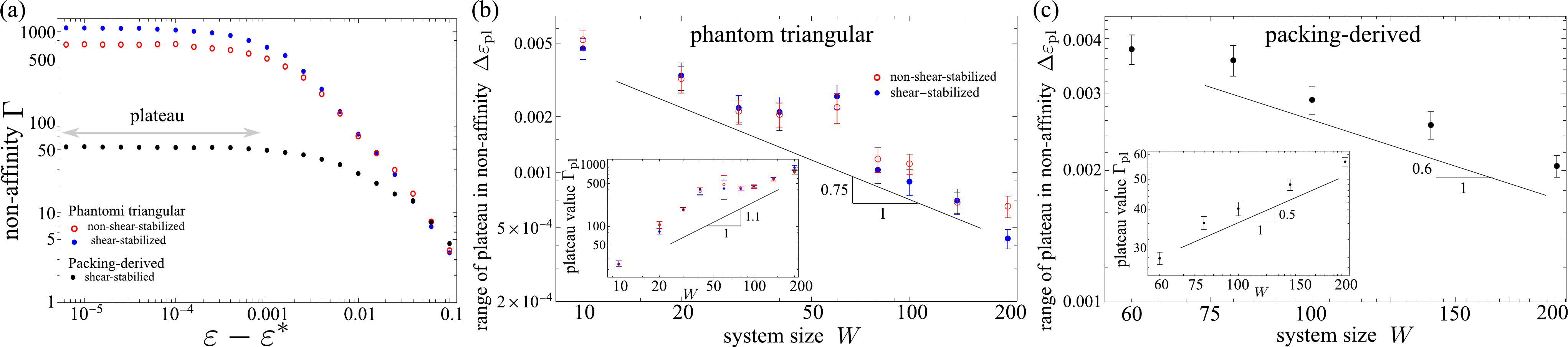}
	\caption{Finite-size scaling of the non-affinity parameter $\Gamma$ defined in Eq.~\eqref{eq:definition_gamma}.
	(a) A plateau regime is observed in $\Gamma$ at strain values close to the transition point $\varepsilon^\ast$. Here we show only three example networks with $W=140$ of different class.
	(b,c) As system size $W$ increases, the range $\Delta \varepsilon_\mathrm{pl}$ of the plateau decreases as a power-law, while the plateau value $\Gamma_\mathrm{pl}$ increases as a power-law. The plateau range $\Delta \varepsilon_\mathrm{pl}$ here is defined as the strain difference between the transition point $\varepsilon^\ast$ and the point where $\Gamma$ shows a 10\% deviation from its plateau value $\Gamma_\mathrm{pl}$. 
	Error bars indicate the standard error of the mean ($\approx50$ networks for each system size).
	Here we studied the same set of networks as in \autoref{fig:upper bound for linear relation}.
	}
	\label{fig:nonAff}
\end{figure*}

\section{Finite-size effects}
\label{sec:finite-size effects}
In the main text, we examined the range of validity of the linear scaling of $\bar{\ell}_\mathrm{min}$ with $\sigma_\ell$.
In \autoref{fig:defEpsimax} we show how we determined this range using a 10\% cutoff on the ratio $\sigma_{\ell}/(\bar{\ell}-\ell_0)$.
We find in \autoref{sec:not from finite-size effect} that this range does not or only weakly change with system size.

Strain-stiffening of spring networks has been shown to be a critical transition when using shear strain as control parameter, which includes diverging fluctuations when approaching the transition point \cite{Sharma2016a,Shivers2019,Arzash2019,Arzash2020}. We wondered whether we would also observe diverging fluctuations when using isotropic strain as control parameter.
Analogous to previous work \cite{Sharma2016a,Shivers2019,Arzash2019,Arzash2020}, we quantify fluctuations using a non-affinity parameter $\Gamma$, which we define as:
\begin{equation}
  \Gamma = \frac{\sum_n{(\delta\bm{u}_n^\mathrm{NA})^2}}{N L_c^2\delta\varepsilon^2}.\label{eq:definition_gamma}
\end{equation}
Here, $\delta\bm{u}_n^\mathrm{NA}$ is the (dimensionful) non-affine displacement of node $n$ during an isotropic expansion by strain $\delta\varepsilon$.
The factor $L_c$ is the length unit defined below Eq.~\eqref{eq:E_general}. 
Because the affine transformation corresponds in our case to uniform isotropic inflation,  Eq.~\eqref{eq:definition_gamma} can be simplified using dimensionless node positions $\bm{r}$, so that in practise we compute $\Gamma$ as:
\begin{equation}\label{eq:nonAff def}
\Gamma(\varepsilon_k) =  \frac{\left(1+\varepsilon_k\right)^2\sum_n\left(\bm{r}_{n,k}-\bm{r}_{n,k-1} \right)^2}{N\left(\varepsilon_k-\varepsilon_{k-1}\right)^2}.
\end{equation}
Here, $\varepsilon_k$ is the strain step with index $k$ within a sweep, and $\bm{r}_{n,k}$ is the corresponding dimensionless position of node $n$.

Using the same networks as in \autoref{fig:upper bound for linear relation}, we numerically studied $\Gamma(\varepsilon)$ and its dependence on system size. 
For all system sizes we observed a plateau in $\Gamma(\varepsilon)$ (\autoref{fig:nonAff}a), for both phantom triangular and packing-derived networks.

We examined how both height $\Gamma_\mathrm{pl}$ and extent $\Delta\varepsilon_\mathrm{pl}$ of the plateau depend on system size (\autoref{fig:nonAff}b,c). We quantified the extent $\Delta\varepsilon_\mathrm{pl}$ as the value of $\varepsilon$ where $\Gamma$ deviates by 10\% from the plateau value, where we also performed linear interpolation between neighboring $\varepsilon_k$ values. 
We found power law scaling with system size of both plateau value $\Gamma_\mathrm{pl} \sim W^{\lambda_\Gamma/\nu_\Gamma}$ and plateau extent $\Delta \varepsilon_\mathrm{pl}\sim W^{-1/\nu_\Gamma}$.
For the phantom triangular networks we found the plateau height exponent $\lambda_\Gamma/\nu_\Gamma\approx 1.1$ and for plateau extent $1/\nu_\Gamma\approx 0.75$.
For the packing-derived networks, we found the plateau height exponent $\lambda_\Gamma/\nu_\Gamma\approx 0.5$ and for plateau extent $1/\nu_\Gamma\approx 0.6$.

These findings are consistent with the finite-size scaling behavior of the non-affinity parameter $\Gamma$ with respect to shear strain $\gamma$ \cite{Sharma2016a,Shivers2019,Arzash2019,Arzash2020}:
The non-affinity parameter generally diverges when approaching the transition as $\Gamma\sim \vert\Delta\gamma\vert^{-\lambda_\Gamma}$ with $\lambda_\Gamma>0$, but
a diverging length scale $\xi_\Gamma \sim \vert\Delta\gamma\vert^{-\nu_\Gamma}$
with $\nu_\Gamma>0$ changes this behavior for system sizes $W\lesssim \xi_\Gamma$.
As a consequence, the non-affinity parameter $\Gamma(\gamma)$ has a plateau whose height scales as $\Gamma_\mathrm{pl}\sim W^{\lambda_\Gamma/\nu_\Gamma}$, and whose extent scales as $\Delta\gamma_\mathrm{pl}\sim W^{-1/\nu_\Gamma}$.
Here we demonstrated that this behavior also appears when using isotropic strain instead of shear strain as control parameter.

Noticeably, the $1/\nu_\Gamma$ values are much larger than the power-law exponent $1/\nu_\ell$ that characterizes the range of the linear scaling of $\bar\ell_\mathrm{min}$ (\autoref{fig:upper bound for linear relation}; for phantom triangular networks $1/\nu_\ell\in [0.04, 0.35]$, while for packing-derived ones $1/\nu_\ell\in [-0.03, 0.08]$). This indicates that the linear scaling regime of $\bar\ell_\mathrm{min}$ is not controlled by the diverging length scale $\xi_\Gamma$ that governs the apparent divergence of the non-affine motions.

\section{General form of the minimal length function} \label{sec:lmin general}

In Eq.~\eqref{eq:Lmin expansion}, we Taylor-expanded the minimal length function $\bar{\ell}_\mathrm{min}$ to the second order in shear strain $\gamma$, while treating the coefficient $a_\ell$ as independent of $\gamma$. Here we discuss a more general form of $\bar{\ell}_\mathrm{min}$ that can include potentially non-analytic dependencies on $\gamma$: 
\begin{equation} \label{eq:lmin general}
\bar{\ell}_\mathrm{min}=\ell_0^\ast -a_\ell(\gamma) \sigma_\ell + g(\gamma) , 
\end{equation}
where the coefficient $a_\ell(\gamma)$ is a function of $\gamma$. We also newly introduced the function $g(\gamma)$, where we choose the convention $g(0)=0$; any offset can be absorbed into $\ell_0^\ast$.  Note that Eq.~\eqref{eq:lmin general} reflects an arbitrary dependency of $\bar{\ell}_\mathrm{min}$ on $\gamma$, while we keep the linear dependency on $\sigma_\ell$.

After minimizing the energy with respect to inner degrees of freedom and the standard deviation $\sigma_\ell$, the resultant energy $e$ is:
\begin{equation}
e = \frac{N}{1+a_\ell^2}\left( \Delta\ell + g\right)^2
\end{equation}
Here we defined $\Delta\ell:=\ell_0^\ast-\ell_0$.
Using $G=(\mathrm{d}^2 e/\mathrm{d}\gamma^2)\vert_{\gamma=0}/N$ we then obtain for the shear modulus:
\begin{equation}
G=\frac{2}{\left(1+a_\ell^2\right)^3}\Big[
Q_0+Q_1\Delta\ell+ Q_2\Delta\ell^2
\Big],\label{eq:G general}
\end{equation}
where $Q_0$, $Q_1$, and $Q_2$ are coefficients that depend only on $a_\ell$, $g$, and their derivatives with respect to $\gamma$.

Note that according to Eq.~\eqref{eq:P prediction}, $\Delta\ell$ is proportional to isotropic tension: $T\sim\Delta\ell$, which can be understood as a consequence of the bulk modulus discontinuity.
Since the $Q$s in Eq.~\eqref{eq:G general} do not depend on $\Delta\ell$, one already observes from this equation that any scaling $G\sim T^\alpha$ needs to have $\alpha\in\lbrace0,1,2\rbrace$. In this sense, the integer scaling exponent between $G$ and $T$ is inherited from the linear scaling of the $\bar\ell_\mathrm{min}$ function with $\sigma_\ell$ in Eq.~\eqref{eq:lmin general}.

Which of the three values $\alpha\in\lbrace0,1,2\rbrace$ is attained depends on the coefficients $Q_0$ to $Q_2$ in Eq.~\eqref{eq:G general}, which are:
\begin{align}
Q_0&= \left(1+a_\ell^2\right)^2{g'}^2,\label{eq:Q0}\\
Q_1&= \left(1+a_\ell^2\right)\Big[\left(1+a_\ell^2\right)g'' -4 a_\ell a_\ell' g' \Big],\label{eq:Q1}\\
Q_2&= a_\ell\left(1+a_\ell^2\right)a_\ell'' + \left(1-3 a_\ell^2\right){a_\ell'}^2.\label{eq:Q2}
\end{align}
Here, for simplicity we used the superscripts ${}'$ and ${}''$ for the first and second derivatives with respect to $\gamma$, respectively.
From Eqs.~\eqref{eq:G general}--\eqref{eq:Q2} follows that the system has finite shear modulus only if at least one of $a_\ell$ or $g$ has a finite first or second derivative with respect to $\gamma$ at $\gamma=0$.

With respect to the scaling exponent $\alpha$ we can say that first, $\alpha=0$ only if $g$ has a finite first derivative. This corresponds to the case where there is a discontinuity in the shear modulus at the transition point. In these cases, the SSS that appears at the transition must have finite overlap with the shear deformation, i.e.\ the network is asymmetric (\autoref{sec:effect of shear strain}, e.g.\ the non-shear stabilized simulations in \autoref{fig:G_P}).
Second, $\alpha=1$ only if $g'=0$ (i.e.\ the network is symmetric) and $g$ has a finite second derivative. This is the typical case that we observe for shear-stabilized networks.
Third, $\alpha=2$ would appear if the network is symmetric ($g'=0$), the second derivative of $g$ vanishes, and $a_\ell$ has finite first or second derivative. This situation might appear at a bifurcation (where $g''$ as bifurcation parameter crosses zero), possibly related to a structural transition in the network.

\end{appendix} 

\bibliographystyle{apsrev4-1}
\bibliography{references}

\begin{thebibliography}{37}%
\makeatletter
\providecommand \@ifxundefined [1]{%
 \@ifx{#1\undefined}
}%
\providecommand \@ifnum [1]{%
 \ifnum #1\expandafter \@firstoftwo
 \else \expandafter \@secondoftwo
 \fi
}%
\providecommand \@ifx [1]{%
 \ifx #1\expandafter \@firstoftwo
 \else \expandafter \@secondoftwo
 \fi
}%
\providecommand \natexlab [1]{#1}%
\providecommand \enquote  [1]{``#1''}%
\providecommand \bibnamefont  [1]{#1}%
\providecommand \bibfnamefont [1]{#1}%
\providecommand \citenamefont [1]{#1}%
\providecommand \href@noop [0]{\@secondoftwo}%
\providecommand \href [0]{\begingroup \@sanitize@url \@href}%
\providecommand \@href[1]{\@@startlink{#1}\@@href}%
\providecommand \@@href[1]{\endgroup#1\@@endlink}%
\providecommand \@sanitize@url [0]{\catcode `\\12\catcode `\$12\catcode
  `\&12\catcode `\#12\catcode `\^12\catcode `\_12\catcode `\%12\relax}%
\providecommand \@@startlink[1]{}%
\providecommand \@@endlink[0]{}%
\providecommand \url  [0]{\begingroup\@sanitize@url \@url }%
\providecommand \@url [1]{\endgroup\@href {#1}{\urlprefix }}%
\providecommand \urlprefix  [0]{URL }%
\providecommand \Eprint [0]{\href }%
\providecommand \doibase [0]{http://dx.doi.org/}%
\providecommand \selectlanguage [0]{\@gobble}%
\providecommand \bibinfo  [0]{\@secondoftwo}%
\providecommand \bibfield  [0]{\@secondoftwo}%
\providecommand \translation [1]{[#1]}%
\providecommand \BibitemOpen [0]{}%
\providecommand \bibitemStop [0]{}%
\providecommand \bibitemNoStop [0]{.\EOS\space}%
\providecommand \EOS [0]{\spacefactor3000\relax}%
\providecommand \BibitemShut  [1]{\csname bibitem#1\endcsname}%
\let\auto@bib@innerbib\@empty
\bibitem [{\citenamefont {Ellenbroek}\ \emph {et~al.}(2009)\citenamefont
  {Ellenbroek}, \citenamefont {Zeravcic}, \citenamefont {{Van Saarloos}},\ and\
  \citenamefont {{Van Hecke}}}]{Ellenbroek2009c}%
  \BibitemOpen
  \bibfield  {author} {\bibinfo {author} {\bibfnamefont {W.~G.}\ \bibnamefont
  {Ellenbroek}}, \bibinfo {author} {\bibfnamefont {Z.}~\bibnamefont
  {Zeravcic}}, \bibinfo {author} {\bibfnamefont {W.}~\bibnamefont {{Van
  Saarloos}}}, \ and\ \bibinfo {author} {\bibfnamefont {M.}~\bibnamefont {{Van
  Hecke}}},\ }\href {\doibase 10.1209/0295-5075/87/34004} {\bibfield  {journal}
  {\bibinfo  {journal} {Epl}\ }\textbf {\bibinfo {volume} {87}},\ \bibinfo
  {pages} {0} (\bibinfo {year} {2009})},\ \Eprint
  {http://arxiv.org/abs/0907.0012} {arXiv:0907.0012} \BibitemShut {NoStop}%
\bibitem [{\citenamefont {Silverberg}\ \emph {et~al.}(2014)\citenamefont
  {Silverberg}, \citenamefont {Barrett}, \citenamefont {Das}, \citenamefont
  {Petersen}, \citenamefont {Bonassar},\ and\ \citenamefont
  {Cohen}}]{Silverberg2014}%
  \BibitemOpen
  \bibfield  {author} {\bibinfo {author} {\bibfnamefont {J.~L.}\ \bibnamefont
  {Silverberg}}, \bibinfo {author} {\bibfnamefont {A.~R.}\ \bibnamefont
  {Barrett}}, \bibinfo {author} {\bibfnamefont {M.}~\bibnamefont {Das}},
  \bibinfo {author} {\bibfnamefont {P.~B.}\ \bibnamefont {Petersen}}, \bibinfo
  {author} {\bibfnamefont {L.~J.}\ \bibnamefont {Bonassar}}, \ and\ \bibinfo
  {author} {\bibfnamefont {I.}~\bibnamefont {Cohen}},\ }\href {\doibase
  10.1016/j.bpj.2014.08.011} {\bibfield  {journal} {\bibinfo  {journal}
  {Biophysical Journal}\ }\textbf {\bibinfo {volume} {107}},\ \bibinfo {pages}
  {1721} (\bibinfo {year} {2014})}\BibitemShut {NoStop}%
\bibitem [{\citenamefont {Licup}\ \emph {et~al.}(2015)\citenamefont {Licup},
  \citenamefont {Münster}, \citenamefont {Sharma}, \citenamefont {Sheinman},
  \citenamefont {Jawerth}, \citenamefont {Fabry}, \citenamefont {Weitz},\ and\
  \citenamefont {MacKintosh}}]{Licup2015}%
  \BibitemOpen
  \bibfield  {author} {\bibinfo {author} {\bibfnamefont {A.~J.}\ \bibnamefont
  {Licup}}, \bibinfo {author} {\bibfnamefont {S.}~\bibnamefont {Münster}},
  \bibinfo {author} {\bibfnamefont {A.}~\bibnamefont {Sharma}}, \bibinfo
  {author} {\bibfnamefont {M.}~\bibnamefont {Sheinman}}, \bibinfo {author}
  {\bibfnamefont {L.~M.}\ \bibnamefont {Jawerth}}, \bibinfo {author}
  {\bibfnamefont {B.}~\bibnamefont {Fabry}}, \bibinfo {author} {\bibfnamefont
  {D.~A.}\ \bibnamefont {Weitz}}, \ and\ \bibinfo {author} {\bibfnamefont
  {F.~C.}\ \bibnamefont {MacKintosh}},\ }\href {\doibase
  10.1073/pnas.1504258112} {\bibfield  {journal} {\bibinfo  {journal}
  {Proceedings of the National Academy of Sciences of the United States of
  America}\ }\textbf {\bibinfo {volume} {112}},\ \bibinfo {pages} {9573}
  (\bibinfo {year} {2015})}\BibitemShut {NoStop}%
\bibitem [{\citenamefont {Feng}\ \emph {et~al.}(2016)\citenamefont {Feng},
  \citenamefont {Levine}, \citenamefont {Mao},\ and\ \citenamefont
  {Sander}}]{Feng2016}%
  \BibitemOpen
  \bibfield  {author} {\bibinfo {author} {\bibfnamefont {J.}~\bibnamefont
  {Feng}}, \bibinfo {author} {\bibfnamefont {H.}~\bibnamefont {Levine}},
  \bibinfo {author} {\bibfnamefont {X.}~\bibnamefont {Mao}}, \ and\ \bibinfo
  {author} {\bibfnamefont {L.~M.}\ \bibnamefont {Sander}},\ }\href {\doibase
  10.1039/c5sm01856k} {\bibfield  {journal} {\bibinfo  {journal} {Soft Matter}\
  }\textbf {\bibinfo {volume} {12}},\ \bibinfo {pages} {1419} (\bibinfo {year}
  {2016})}\BibitemShut {NoStop}%
\bibitem [{\citenamefont {Oosten}\ \emph {et~al.}(2016)\citenamefont {Oosten},
  \citenamefont {Vahabi}, \citenamefont {Licup}, \citenamefont {Sharma},
  \citenamefont {Galie}, \citenamefont {MacKintosh},\ and\ \citenamefont
  {Janmey}}]{Van_Oosten2016}%
  \BibitemOpen
  \bibfield  {author} {\bibinfo {author} {\bibfnamefont {A.~S.~V.}\
  \bibnamefont {Oosten}}, \bibinfo {author} {\bibfnamefont {M.}~\bibnamefont
  {Vahabi}}, \bibinfo {author} {\bibfnamefont {A.~J.}\ \bibnamefont {Licup}},
  \bibinfo {author} {\bibfnamefont {A.}~\bibnamefont {Sharma}}, \bibinfo
  {author} {\bibfnamefont {P.~A.}\ \bibnamefont {Galie}}, \bibinfo {author}
  {\bibfnamefont {F.~C.}\ \bibnamefont {MacKintosh}}, \ and\ \bibinfo {author}
  {\bibfnamefont {P.~A.}\ \bibnamefont {Janmey}},\ }\href {\doibase
  10.1038/srep19270} {\bibfield  {journal} {\bibinfo  {journal} {Scientific
  Reports}\ }\textbf {\bibinfo {volume} {6}} (\bibinfo {year} {2016}),\
  10.1038/srep19270}\BibitemShut {NoStop}%
\bibitem [{\citenamefont {Sharma}\ \emph
  {et~al.}(2016{\natexlab{a}})\citenamefont {Sharma}, \citenamefont {Licup},
  \citenamefont {Rens}, \citenamefont {Vahabi}, \citenamefont {Jansen},
  \citenamefont {Koenderink},\ and\ \citenamefont {MacKintosh}}]{Sharma2016a}%
  \BibitemOpen
  \bibfield  {author} {\bibinfo {author} {\bibfnamefont {A.}~\bibnamefont
  {Sharma}}, \bibinfo {author} {\bibfnamefont {A.~J.}\ \bibnamefont {Licup}},
  \bibinfo {author} {\bibfnamefont {R.}~\bibnamefont {Rens}}, \bibinfo {author}
  {\bibfnamefont {M.}~\bibnamefont {Vahabi}}, \bibinfo {author} {\bibfnamefont
  {K.~A.}\ \bibnamefont {Jansen}}, \bibinfo {author} {\bibfnamefont {G.~H.}\
  \bibnamefont {Koenderink}}, \ and\ \bibinfo {author} {\bibfnamefont {F.~C.}\
  \bibnamefont {MacKintosh}},\ }\href {\doibase 10.1103/PhysRevE.94.042407}
  {\bibfield  {journal} {\bibinfo  {journal} {Physical Review E}\ }\textbf
  {\bibinfo {volume} {94}} (\bibinfo {year} {2016}{\natexlab{a}}),\
  10.1103/PhysRevE.94.042407}\BibitemShut {NoStop}%
\bibitem [{\citenamefont {Sharma}\ \emph
  {et~al.}(2016{\natexlab{b}})\citenamefont {Sharma}, \citenamefont {Licup},
  \citenamefont {Jansen}, \citenamefont {Rens}, \citenamefont {Sheinman},
  \citenamefont {Koenderink},\ and\ \citenamefont {Mackintosh}}]{Sharma2016b}%
  \BibitemOpen
  \bibfield  {author} {\bibinfo {author} {\bibfnamefont {A.}~\bibnamefont
  {Sharma}}, \bibinfo {author} {\bibfnamefont {A.~J.}\ \bibnamefont {Licup}},
  \bibinfo {author} {\bibfnamefont {K.~A.}\ \bibnamefont {Jansen}}, \bibinfo
  {author} {\bibfnamefont {R.}~\bibnamefont {Rens}}, \bibinfo {author}
  {\bibfnamefont {M.}~\bibnamefont {Sheinman}}, \bibinfo {author}
  {\bibfnamefont {G.~H.}\ \bibnamefont {Koenderink}}, \ and\ \bibinfo {author}
  {\bibfnamefont {F.~C.}\ \bibnamefont {Mackintosh}},\ }\href {\doibase
  10.1038/nphys3628} {\bibfield  {journal} {\bibinfo  {journal} {Nature
  Physics}\ }\textbf {\bibinfo {volume} {12}},\ \bibinfo {pages} {584}
  (\bibinfo {year} {2016}{\natexlab{b}})}\BibitemShut {NoStop}%
\bibitem [{\citenamefont {Licup}\ \emph {et~al.}(2016)\citenamefont {Licup},
  \citenamefont {Sharma},\ and\ \citenamefont {Mackintosh}}]{Licup2016}%
  \BibitemOpen
  \bibfield  {author} {\bibinfo {author} {\bibfnamefont {A.~J.}\ \bibnamefont
  {Licup}}, \bibinfo {author} {\bibfnamefont {A.}~\bibnamefont {Sharma}}, \
  and\ \bibinfo {author} {\bibfnamefont {F.~C.}\ \bibnamefont {Mackintosh}},\
  }\href {\doibase 10.1103/PhysRevE.93.012407} {\bibfield  {journal} {\bibinfo
  {journal} {Physical Review E}\ }\textbf {\bibinfo {volume} {93}} (\bibinfo
  {year} {2016}),\ 10.1103/PhysRevE.93.012407}\BibitemShut {NoStop}%
\bibitem [{\citenamefont {Jansen}\ \emph {et~al.}(2018)\citenamefont {Jansen},
  \citenamefont {Licup}, \citenamefont {Sharma}, \citenamefont {Rens},
  \citenamefont {MacKintosh},\ and\ \citenamefont {Koenderink}}]{Jansen2018}%
  \BibitemOpen
  \bibfield  {author} {\bibinfo {author} {\bibfnamefont {K.~A.}\ \bibnamefont
  {Jansen}}, \bibinfo {author} {\bibfnamefont {A.~J.}\ \bibnamefont {Licup}},
  \bibinfo {author} {\bibfnamefont {A.}~\bibnamefont {Sharma}}, \bibinfo
  {author} {\bibfnamefont {R.}~\bibnamefont {Rens}}, \bibinfo {author}
  {\bibfnamefont {F.~C.}\ \bibnamefont {MacKintosh}}, \ and\ \bibinfo {author}
  {\bibfnamefont {G.~H.}\ \bibnamefont {Koenderink}},\ }\href {\doibase
  10.1016/j.bpj.2018.04.043} {\bibfield  {journal} {\bibinfo  {journal}
  {Biophysical Journal}\ }\textbf {\bibinfo {volume} {114}},\ \bibinfo {pages}
  {2665} (\bibinfo {year} {2018})}\BibitemShut {NoStop}%
\bibitem [{\citenamefont {Shivers}\ \emph {et~al.}(2019)\citenamefont
  {Shivers}, \citenamefont {Arzash}, \citenamefont {Sharma},\ and\
  \citenamefont {MacKintosh}}]{Shivers2019}%
  \BibitemOpen
  \bibfield  {author} {\bibinfo {author} {\bibfnamefont {J.~L.}\ \bibnamefont
  {Shivers}}, \bibinfo {author} {\bibfnamefont {S.}~\bibnamefont {Arzash}},
  \bibinfo {author} {\bibfnamefont {A.}~\bibnamefont {Sharma}}, \ and\ \bibinfo
  {author} {\bibfnamefont {F.~C.}\ \bibnamefont {MacKintosh}},\ }\href
  {\doibase 10.1103/PhysRevLett.122.188003} {\bibfield  {journal} {\bibinfo
  {journal} {Physical Review Letters}\ }\textbf {\bibinfo {volume} {122}}
  (\bibinfo {year} {2019}),\ 10.1103/PhysRevLett.122.188003}\BibitemShut
  {NoStop}%
\bibitem [{\citenamefont {Arzash}\ \emph {et~al.}(2020)\citenamefont {Arzash},
  \citenamefont {Shivers},\ and\ \citenamefont {MacKintosh}}]{Arzash2020}%
  \BibitemOpen
  \bibfield  {author} {\bibinfo {author} {\bibfnamefont {S.}~\bibnamefont
  {Arzash}}, \bibinfo {author} {\bibfnamefont {J.~L.}\ \bibnamefont {Shivers}},
  \ and\ \bibinfo {author} {\bibfnamefont {F.~C.}\ \bibnamefont {MacKintosh}},\
  }\href {\doibase 10.1039/D0SM00764A} {\bibfield  {journal} {\bibinfo
  {journal} {Soft Matter}\ }\textbf {\bibinfo {volume} {169}},\ \bibinfo
  {pages} {6784} (\bibinfo {year} {2020})}\BibitemShut {NoStop}%
\bibitem [{\citenamefont {Maxwell}(1864)}]{Maxwell1864}%
  \BibitemOpen
  \bibfield  {author} {\bibinfo {author} {\bibfnamefont {J.~C.}\ \bibnamefont
  {Maxwell}},\ }\href {\doibase 10.1080/14786446408643668} {\bibfield
  {journal} {\bibinfo  {journal} {The London, Edinburgh, and Dublin
  Philosophical Magazine and Journal of Science}\ }\textbf {\bibinfo {volume}
  {27}},\ \bibinfo {pages} {294} (\bibinfo {year} {1864})}\BibitemShut
  {NoStop}%
\bibitem [{\citenamefont {Calladine}(1978)}]{Calladine1978}%
  \BibitemOpen
  \bibfield  {author} {\bibinfo {author} {\bibfnamefont {C.}~\bibnamefont
  {Calladine}},\ }\href {\doibase https://doi.org/10.1016/0020-7683(78)90052-5}
  {\bibfield  {journal} {\bibinfo  {journal} {International Journal of Solids
  and Structures}\ }\textbf {\bibinfo {volume} {14}},\ \bibinfo {pages} {161}
  (\bibinfo {year} {1978})}\BibitemShut {NoStop}%
\bibitem [{\citenamefont {Lubensky}\ \emph {et~al.}(2015)\citenamefont
  {Lubensky}, \citenamefont {Kane}, \citenamefont {Mao}, \citenamefont
  {Souslov},\ and\ \citenamefont {Sun}}]{Lubensky2015}%
  \BibitemOpen
  \bibfield  {author} {\bibinfo {author} {\bibfnamefont {T.~C.}\ \bibnamefont
  {Lubensky}}, \bibinfo {author} {\bibfnamefont {C.~L.}\ \bibnamefont {Kane}},
  \bibinfo {author} {\bibfnamefont {X.}~\bibnamefont {Mao}}, \bibinfo {author}
  {\bibfnamefont {A.}~\bibnamefont {Souslov}}, \ and\ \bibinfo {author}
  {\bibfnamefont {K.}~\bibnamefont {Sun}},\ }\href {\doibase
  10.1088/0034-4885/78/7/073901} {\bibfield  {journal} {\bibinfo  {journal}
  {Reports Prog. Phys.}\ }\textbf {\bibinfo {volume} {78}},\ \bibinfo {pages}
  {73901} (\bibinfo {year} {2015})},\ \Eprint {http://arxiv.org/abs/1503.01324}
  {1503.01324} \BibitemShut {NoStop}%
\bibitem [{\citenamefont {Merkel}\ \emph {et~al.}(2019)\citenamefont {Merkel},
  \citenamefont {Baumgarten}, \citenamefont {Tighe},\ and\ \citenamefont
  {Manning}}]{Merkel2019}%
  \BibitemOpen
  \bibfield  {author} {\bibinfo {author} {\bibfnamefont {M.}~\bibnamefont
  {Merkel}}, \bibinfo {author} {\bibfnamefont {K.}~\bibnamefont {Baumgarten}},
  \bibinfo {author} {\bibfnamefont {B.~P.}\ \bibnamefont {Tighe}}, \ and\
  \bibinfo {author} {\bibfnamefont {M.~L.}\ \bibnamefont {Manning}},\ }\href
  {\doibase 10.1073/pnas.1815436116} {\bibfield  {journal} {\bibinfo  {journal}
  {Proc. Natl. Acad. Sci.}\ }\textbf {\bibinfo {volume} {116}},\ \bibinfo
  {pages} {6560} (\bibinfo {year} {2019})}\BibitemShut {NoStop}%
\bibitem [{\citenamefont {Alexander}(1998)}]{Alexander1998}%
  \BibitemOpen
  \bibfield  {author} {\bibinfo {author} {\bibfnamefont {S.}~\bibnamefont
  {Alexander}},\ }\href {\doibase
  https://doi.org/10.1016/S0370-1573(97)00069-0} {\bibfield  {journal}
  {\bibinfo  {journal} {Physics Reports}\ }\textbf {\bibinfo {volume} {296}},\
  \bibinfo {pages} {65} (\bibinfo {year} {1998})}\BibitemShut {NoStop}%
\bibitem [{\citenamefont {Wyart}\ \emph {et~al.}(2008)\citenamefont {Wyart},
  \citenamefont {Liang}, \citenamefont {Kabla},\ and\ \citenamefont
  {Mahadevan}}]{Wyart2008}%
  \BibitemOpen
  \bibfield  {author} {\bibinfo {author} {\bibfnamefont {M.}~\bibnamefont
  {Wyart}}, \bibinfo {author} {\bibfnamefont {H.}~\bibnamefont {Liang}},
  \bibinfo {author} {\bibfnamefont {A.}~\bibnamefont {Kabla}}, \ and\ \bibinfo
  {author} {\bibfnamefont {L.}~\bibnamefont {Mahadevan}},\ }\href {\doibase
  10.1103/PhysRevLett.101.215501} {\bibfield  {journal} {\bibinfo  {journal}
  {Phys. Rev. Lett.}\ }\textbf {\bibinfo {volume} {101}},\ \bibinfo {pages} {1}
  (\bibinfo {year} {2008})},\ \Eprint {http://arxiv.org/abs/0806.4571v1}
  {arXiv:0806.4571v1} \BibitemShut {NoStop}%
\bibitem [{\citenamefont {Ingber}\ \emph {et~al.}(2014)\citenamefont {Ingber},
  \citenamefont {Wang},\ and\ \citenamefont {Stamenovi{\'{c}}}}]{Ingber2014}%
  \BibitemOpen
  \bibfield  {author} {\bibinfo {author} {\bibfnamefont {D.~E.}\ \bibnamefont
  {Ingber}}, \bibinfo {author} {\bibfnamefont {N.}~\bibnamefont {Wang}}, \ and\
  \bibinfo {author} {\bibfnamefont {D.}~\bibnamefont {Stamenovi{\'{c}}}},\
  }\href {\doibase 10.1088/0034-4885/77/4/046603} {\bibfield  {journal}
  {\bibinfo  {journal} {Reports on Progress in Physics}\ }\textbf {\bibinfo
  {volume} {77}},\ \bibinfo {pages} {046603} (\bibinfo {year}
  {2014})}\BibitemShut {NoStop}%
\bibitem [{\citenamefont {Arzash}\ \emph {et~al.}(2019)\citenamefont {Arzash},
  \citenamefont {Shivers}, \citenamefont {Licup}, \citenamefont {Sharma},\ and\
  \citenamefont {MacKintosh}}]{Arzash2019}%
  \BibitemOpen
  \bibfield  {author} {\bibinfo {author} {\bibfnamefont {S.}~\bibnamefont
  {Arzash}}, \bibinfo {author} {\bibfnamefont {J.~L.}\ \bibnamefont {Shivers}},
  \bibinfo {author} {\bibfnamefont {A.~J.}\ \bibnamefont {Licup}}, \bibinfo
  {author} {\bibfnamefont {A.}~\bibnamefont {Sharma}}, \ and\ \bibinfo {author}
  {\bibfnamefont {F.~C.}\ \bibnamefont {MacKintosh}},\ }\href {\doibase
  10.1103/PhysRevE.99.042412} {\bibfield  {journal} {\bibinfo  {journal} {Phys.
  Rev. E}\ }\textbf {\bibinfo {volume} {99}},\ \bibinfo {pages} {042412}
  (\bibinfo {year} {2019})}\BibitemShut {NoStop}%
\bibitem [{\citenamefont {Cui}\ \emph {et~al.}(2019)\citenamefont {Cui},
  \citenamefont {Ruocco},\ and\ \citenamefont {Zaccone}}]{Cui2019}%
  \BibitemOpen
  \bibfield  {author} {\bibinfo {author} {\bibfnamefont {B.}~\bibnamefont
  {Cui}}, \bibinfo {author} {\bibfnamefont {G.}~\bibnamefont {Ruocco}}, \ and\
  \bibinfo {author} {\bibfnamefont {A.}~\bibnamefont {Zaccone}},\ }\href
  {\doibase 10.1007/s10035-019-0916-4} {\bibfield  {journal} {\bibinfo
  {journal} {Granular Matter}\ }\textbf {\bibinfo {volume} {21}} (\bibinfo
  {year} {2019}),\ 10.1007/s10035-019-0916-4}\BibitemShut {NoStop}%
\bibitem [{\citenamefont {Damavandi}\ \emph
  {et~al.}(2021{\natexlab{a}})\citenamefont {Damavandi}, \citenamefont {Hagh},
  \citenamefont {Santangelo},\ and\ \citenamefont {Manning}}]{Damavandi2021}%
  \BibitemOpen
  \bibfield  {author} {\bibinfo {author} {\bibfnamefont {O.~K.}\ \bibnamefont
  {Damavandi}}, \bibinfo {author} {\bibfnamefont {V.~F.}\ \bibnamefont {Hagh}},
  \bibinfo {author} {\bibfnamefont {C.~D.}\ \bibnamefont {Santangelo}}, \ and\
  \bibinfo {author} {\bibfnamefont {M.~L.}\ \bibnamefont {Manning}},\ }\href
  {http://arxiv.org/abs/2102.11310} {\bibfield  {journal} {\bibinfo  {journal}
  {arXiv}\ } (\bibinfo {year} {2021}{\natexlab{a}})},\ \Eprint
  {http://arxiv.org/abs/2102.11310} {2102.11310} \BibitemShut {NoStop}%
\bibitem [{\citenamefont {Onck}\ \emph {et~al.}(2005)\citenamefont {Onck},
  \citenamefont {Koeman}, \citenamefont {van Dillen},\ and\ \citenamefont
  {van~der Giessen}}]{Onck2005}%
  \BibitemOpen
  \bibfield  {author} {\bibinfo {author} {\bibfnamefont {P.~R.}\ \bibnamefont
  {Onck}}, \bibinfo {author} {\bibfnamefont {T.}~\bibnamefont {Koeman}},
  \bibinfo {author} {\bibfnamefont {T.}~\bibnamefont {van Dillen}}, \ and\
  \bibinfo {author} {\bibfnamefont {E.}~\bibnamefont {van~der Giessen}},\
  }\href {\doibase 10.1103/PhysRevLett.95.178102} {\bibfield  {journal}
  {\bibinfo  {journal} {Phys. Rev. Lett.}\ }\textbf {\bibinfo {volume} {95}},\
  \bibinfo {pages} {178102} (\bibinfo {year} {2005})}\BibitemShut {NoStop}%
\bibitem [{\citenamefont {Sheinman}\ \emph {et~al.}(2012)\citenamefont
  {Sheinman}, \citenamefont {Broedersz},\ and\ \citenamefont
  {MacKintosh}}]{Sheinman2012}%
  \BibitemOpen
  \bibfield  {author} {\bibinfo {author} {\bibfnamefont {M.}~\bibnamefont
  {Sheinman}}, \bibinfo {author} {\bibfnamefont {C.~P.}\ \bibnamefont
  {Broedersz}}, \ and\ \bibinfo {author} {\bibfnamefont {F.~C.}\ \bibnamefont
  {MacKintosh}},\ }\href {\doibase 10.1103/PhysRevE.85.021801} {\bibfield
  {journal} {\bibinfo  {journal} {Physical Review E - Statistical, Nonlinear,
  and Soft Matter Physics}\ }\textbf {\bibinfo {volume} {85}} (\bibinfo {year}
  {2012}),\ 10.1103/PhysRevE.85.021801}\BibitemShut {NoStop}%
\bibitem [{\citenamefont {Vermeulen}\ \emph {et~al.}(2017)\citenamefont
  {Vermeulen}, \citenamefont {Bose}, \citenamefont {Storm},\ and\ \citenamefont
  {Ellenbroek}}]{Vermeulen2017}%
  \BibitemOpen
  \bibfield  {author} {\bibinfo {author} {\bibfnamefont {M.~F.~J.}\
  \bibnamefont {Vermeulen}}, \bibinfo {author} {\bibfnamefont {A.}~\bibnamefont
  {Bose}}, \bibinfo {author} {\bibfnamefont {C.}~\bibnamefont {Storm}}, \ and\
  \bibinfo {author} {\bibfnamefont {W.~G.}\ \bibnamefont {Ellenbroek}},\ }\href
  {\doibase 10.1103/PhysRevE.96.053003} {\bibfield  {journal} {\bibinfo
  {journal} {Phys. Rev. E}\ }\textbf {\bibinfo {volume} {96}},\ \bibinfo
  {pages} {053003} (\bibinfo {year} {2017})}\BibitemShut {NoStop}%
\bibitem [{\citenamefont {D{\"{u}}ring}\ \emph {et~al.}(2014)\citenamefont
  {D{\"{u}}ring}, \citenamefont {Lerner},\ and\ \citenamefont
  {Wyart}}]{During2014}%
  \BibitemOpen
  \bibfield  {author} {\bibinfo {author} {\bibfnamefont {G.}~\bibnamefont
  {D{\"{u}}ring}}, \bibinfo {author} {\bibfnamefont {E.}~\bibnamefont
  {Lerner}}, \ and\ \bibinfo {author} {\bibfnamefont {M.}~\bibnamefont
  {Wyart}},\ }\href {\doibase 10.1103/PhysRevE.89.022305} {\bibfield  {journal}
  {\bibinfo  {journal} {Phys. Rev. E - Stat. Nonlinear, Soft Matter Phys.}\
  }\textbf {\bibinfo {volume} {89}},\ \bibinfo {pages} {1} (\bibinfo {year}
  {2014})},\ \Eprint {http://arxiv.org/abs/1308.3886} {arXiv:1308.3886}
  \BibitemShut {NoStop}%
\bibitem [{\citenamefont {Rens}\ \emph {et~al.}(2018)\citenamefont {Rens},
  \citenamefont {Villarroel}, \citenamefont {D{\"{u}}ring},\ and\ \citenamefont
  {Lerner}}]{Rens2018}%
  \BibitemOpen
  \bibfield  {author} {\bibinfo {author} {\bibfnamefont {R.}~\bibnamefont
  {Rens}}, \bibinfo {author} {\bibfnamefont {C.}~\bibnamefont {Villarroel}},
  \bibinfo {author} {\bibfnamefont {G.}~\bibnamefont {D{\"{u}}ring}}, \ and\
  \bibinfo {author} {\bibfnamefont {E.}~\bibnamefont {Lerner}},\ }\href
  {\doibase 10.1103/PhysRevE.98.062411} {\bibfield  {journal} {\bibinfo
  {journal} {Phys. Rev. E}\ }\textbf {\bibinfo {volume} {98}},\ \bibinfo
  {pages} {062411} (\bibinfo {year} {2018})}\BibitemShut {NoStop}%
\bibitem [{\citenamefont {Merkel}\ and\ \citenamefont
  {Manning}(2018)}]{Merkel2018}%
  \BibitemOpen
  \bibfield  {author} {\bibinfo {author} {\bibfnamefont {M.}~\bibnamefont
  {Merkel}}\ and\ \bibinfo {author} {\bibfnamefont {M.~L.}\ \bibnamefont
  {Manning}},\ }\href {\doibase 10.1088/1367-2630/aaaa13} {\bibfield  {journal}
  {\bibinfo  {journal} {New J. Phys.}\ }\textbf {\bibinfo {volume} {20}},\
  \bibinfo {pages} {022002} (\bibinfo {year} {2018})}\BibitemShut {NoStop}%
\bibitem [{\citenamefont {Lerner}(2019)}]{Lerner2019}%
  \BibitemOpen
  \bibfield  {author} {\bibinfo {author} {\bibfnamefont {E.}~\bibnamefont
  {Lerner}},\ }\href {\doibase 10.1016/j.jnoncrysol.2019.119570} {\bibfield
  {journal} {\bibinfo  {journal} {J. Non. Cryst. Solids}\ }\textbf {\bibinfo
  {volume} {522}},\ \bibinfo {pages} {1} (\bibinfo {year} {2019})},\ \Eprint
  {http://arxiv.org/abs/1902.08991} {arXiv:1902.08991} \BibitemShut {NoStop}%
\bibitem [{Note1()}]{Note1}%
  \BibitemOpen
  \bibinfo {note} {This is because Eq.~\protect \textup {\hbox {\mathsurround
  \z@ \protect \normalfont (\ignorespaces \ref {eq:sigma in toy model}\unskip
  \@@italiccorr )}} can be transformed into: \begin {equation} \sigma _L^2 =
  \protect \frac {1}{16}(2u^2+9w^2), \end {equation} where $u=\protect \tilde
  {x}_1+\protect \tilde {x}_2$ and $w=\protect \tilde {x}_1-\protect \tilde
  {x}_2$. This is the equation of an ellipse whose main axes are diagonally
  oriented and scale with $\sigma _\ell $.}\BibitemShut {Stop}%
\bibitem [{Note2()}]{Note2}%
  \BibitemOpen
  \bibinfo {note} {That $\protect \bar {\ell }$ can become arbitrarily large
  for given $\sigma _\ell <1/4$ can be shown explicitly by considering a subset
  of configurations parameterized by two scalars $w$ and $h$ as $\protect \bm
  {r}_1=(-w/2,h)$ and $\protect \bm {r}_2=(w/2,h)$. Then one can show that the
  choice \begin {equation*} w(\sigma _\ell ,h) = -\protect \frac
  {1}{3}(1+8\sigma _\ell ) + \protect \frac {2}{3}\protect \sqrt {(1+2\sigma
  _\ell )^2 + 3h^2} \end {equation*} leads to the correct value for the
  standard deviation of the spring lengths $\sigma _\ell $. Moreover, one can
  show that for this choice, the relation \begin {equation*} \protect \bar
  {\ell }(\sigma _\ell ,h) = \sigma _\ell + w(\sigma _\ell ,h) \end {equation*}
  holds, and that $\protect \bar {\ell }(\sigma _L\ell ,h=0)=\protect \bar
  {\ell }_\protect \mathrm {min}(\sigma _\ell )$ with $\protect \bar {\ell
  }_\protect \mathrm {min}$ given by Eq.~\protect \textup {\hbox {\mathsurround
  \z@ \protect \normalfont (\ignorespaces \ref {eq:linear Lmin eq}\unskip
  \@@italiccorr )}}. Finally, for fixed $\sigma _\ell $, the function $\protect
  \bar {\ell }(\sigma _\ell ,h)$ increases monotonically with $h$ without upper
  bound.}\BibitemShut {Stop}%
\bibitem [{\citenamefont {Broedersz}\ and\ \citenamefont
  {MacKintosh}(2011)}]{Broedersz2011b}%
  \BibitemOpen
  \bibfield  {author} {\bibinfo {author} {\bibfnamefont {C.~P.}\ \bibnamefont
  {Broedersz}}\ and\ \bibinfo {author} {\bibfnamefont {F.~C.}\ \bibnamefont
  {MacKintosh}},\ }\href {\doibase 10.1039/c0sm01004a} {\bibfield  {journal}
  {\bibinfo  {journal} {Soft Matter}\ }\textbf {\bibinfo {volume} {7}},\
  \bibinfo {pages} {3186} (\bibinfo {year} {2011})},\ \Eprint
  {http://arxiv.org/abs/1009.3848} {arXiv:1009.3848} \BibitemShut {NoStop}%
\bibitem [{\citenamefont {Dagois-Bohy}\ \emph {et~al.}(2012)\citenamefont
  {Dagois-Bohy}, \citenamefont {Tighe}, \citenamefont {Simon}, \citenamefont
  {Henkes},\ and\ \citenamefont {{Van Hecke}}}]{Dagois-Bohy2012}%
  \BibitemOpen
  \bibfield  {author} {\bibinfo {author} {\bibfnamefont {S.}~\bibnamefont
  {Dagois-Bohy}}, \bibinfo {author} {\bibfnamefont {B.~P.}\ \bibnamefont
  {Tighe}}, \bibinfo {author} {\bibfnamefont {J.}~\bibnamefont {Simon}},
  \bibinfo {author} {\bibfnamefont {S.}~\bibnamefont {Henkes}}, \ and\ \bibinfo
  {author} {\bibfnamefont {M.}~\bibnamefont {{Van Hecke}}},\ }\href {\doibase
  10.1103/PhysRevLett.109.095703} {\bibfield  {journal} {\bibinfo  {journal}
  {Phys. Rev. Lett.}\ }\textbf {\bibinfo {volume} {109}},\ \bibinfo {pages} {1}
  (\bibinfo {year} {2012})},\ \Eprint {http://arxiv.org/abs/1203.3364}
  {arXiv:1203.3364} \BibitemShut {NoStop}%
\bibitem [{\citenamefont {Damavandi}\ \emph
  {et~al.}(2021{\natexlab{b}})\citenamefont {Damavandi}, \citenamefont
  {Manning},\ and\ \citenamefont {Schwarz}}]{Damavandi2021b}%
  \BibitemOpen
  \bibfield  {author} {\bibinfo {author} {\bibfnamefont {O.~K.}\ \bibnamefont
  {Damavandi}}, \bibinfo {author} {\bibfnamefont {M.~L.}\ \bibnamefont
  {Manning}}, \ and\ \bibinfo {author} {\bibfnamefont {J.~M.}\ \bibnamefont
  {Schwarz}},\ }\href {http://arxiv.org/abs/2110.04343} {\bibfield  {journal}
  {\bibinfo  {journal} {arXiv}\ } (\bibinfo {year} {2021}{\natexlab{b}})},\
  \Eprint {http://arxiv.org/abs/2110.04343} {arXiv:2110.04343} \BibitemShut
  {NoStop}%
\bibitem [{\citenamefont {Lindstr{\"{o}}m}\ \emph {et~al.}(2010)\citenamefont
  {Lindstr{\"{o}}m}, \citenamefont {Vader}, \citenamefont {Kulachenko},\ and\
  \citenamefont {Weitz}}]{Lindstrom2010}%
  \BibitemOpen
  \bibfield  {author} {\bibinfo {author} {\bibfnamefont {S.~B.}\ \bibnamefont
  {Lindstr{\"{o}}m}}, \bibinfo {author} {\bibfnamefont {D.~A.}\ \bibnamefont
  {Vader}}, \bibinfo {author} {\bibfnamefont {A.}~\bibnamefont {Kulachenko}}, \
  and\ \bibinfo {author} {\bibfnamefont {D.~A.}\ \bibnamefont {Weitz}},\ }\href
  {\doibase 10.1103/PhysRevE.82.051905} {\bibfield  {journal} {\bibinfo
  {journal} {Phys. Rev. E}\ }\textbf {\bibinfo {volume} {82}},\ \bibinfo
  {pages} {051905} (\bibinfo {year} {2010})}\BibitemShut {NoStop}%
\bibitem [{\citenamefont {Huang}\ and\ \citenamefont {Born}(1950)}]{Huang1950}%
  \BibitemOpen
  \bibfield  {author} {\bibinfo {author} {\bibfnamefont {K.}~\bibnamefont
  {Huang}}\ and\ \bibinfo {author} {\bibfnamefont {M.}~\bibnamefont {Born}},\
  }\href {\doibase 10.1098/rspa.1950.0133} {\bibfield  {journal} {\bibinfo
  {journal} {Proc. R. Soc. Lond., A Math. phys. sci. A}\ }\textbf {\bibinfo
  {volume} {203}},\ \bibinfo {pages} {178} (\bibinfo {year}
  {1950})}\BibitemShut {NoStop}%
\bibitem [{\citenamefont {Born}\ and\ \citenamefont {Huang}(1955)}]{Born1955}%
  \BibitemOpen
  \bibfield  {author} {\bibinfo {author} {\bibfnamefont {M.}~\bibnamefont
  {Born}}\ and\ \bibinfo {author} {\bibfnamefont {K.}~\bibnamefont {Huang}},\
  }\href {\doibase 10.1119/1.1934059} {\bibfield  {journal} {\bibinfo
  {journal} {Am. J. Phys.}\ }\textbf {\bibinfo {volume} {23}},\ \bibinfo
  {pages} {474} (\bibinfo {year} {1955})},\ \Eprint
  {http://arxiv.org/abs/https://doi.org/10.1119/1.1934059}
  {https://doi.org/10.1119/1.1934059} \BibitemShut {NoStop}%
\bibitem [{\citenamefont {Lema{\^i}tre}\ and\ \citenamefont
  {Maloney}(2006)}]{Lemaitre2006}%
  \BibitemOpen
  \bibfield  {author} {\bibinfo {author} {\bibfnamefont {A.}~\bibnamefont
  {Lema{\^i}tre}}\ and\ \bibinfo {author} {\bibfnamefont {C.}~\bibnamefont
  {Maloney}},\ }\href {\doibase 10.1007/s10955-005-9015-5} {\bibfield
  {journal} {\bibinfo  {journal} {Journal of Statistical Physics}\ }\textbf
  {\bibinfo {volume} {123}},\ \bibinfo {pages} {415} (\bibinfo {year}
  {2006})}\BibitemShut {NoStop}%
\end{thebibliography}%

\end{document}